\documentclass[11pt,letterpaper]{article}
  
  \usepackage[draft]{minted}
  \usepackage[margin=1in]{geometry}
  \usepackage{graphicx}
  \usepackage{caption}
  \usepackage{subcaption}
  \usepackage{amsmath}
  \usepackage{authblk}
  \usepackage{booktabs}
  \usepackage[utf8]{inputenc}
  \usepackage{enumitem}
  \usepackage[numbers]{natbib}
  \usepackage{multirow}
  \usepackage{pdflscape}
  \usepackage{hyperref} % load this last
  
  \newlength{\tindent}
  \DeclareUnicodeCharacter{3F5}{$\epsilon$}
  \setminted[julia]{fontfamily=tt,fontsize=\footnotesize,breaklines}
  
  \let\proglang=\textsf % is this safe?
  \newcommand{\pkg}[1]{{\fontseries{b}\selectfont #1}}
  \newcommand{\code}[1]{\mintinline{julia}{#1}}
  \renewcommand{\mathbf}[1]{\boldsymbol{#1}}
  
  \newtheorem{example}{Example}
  \newcommand{\abs}[1]{\left|#1\right|}
  \newcommand{\ra}[1]{\renewcommand{\arraystretch}{#1}}
  
  \title{
  \vspace{2in}
  \textmd{\textbf{BioSimulator.jl: Stochastic simulation in Julia}}\\
  }
  
  \author[a]{\small Alfonso Landeros}
  \author[a]{\small Timothy Stutz}
  \author[b]{\small Kevin L. Keys}
  \author[c]{\small Alexander Alekseyenko}
  \author[d]{\small Janet S. Sinsheimer}
  \author[a]{\small Kenneth Lange}
  \author[a, $\ast$]{\small Mary Sehl}
  \affil[a]{\small Department of Biomathematics, David Geffen School of Medicine at UCLA, USA}
  \affil[b]{\small Department of Medicine, University of California, San Francisco, USA}
  \affil[c]{\small Department of Public Health Sciences, Medical University of South Carolina, USA}
  \affil[d]{\small Department of Human Genetics, David Geffen School of Medicine at UCLA, USA}
  
  \date{}
\begin{document}

%%%%% title page %%%%%

\maketitle

\vspace{1.5in}

% contact information for corresponding author
\noindent {\scriptsize $\ast$ Corresponding author at: \\
Division of Hematology-Oncology \\
Department of Medicine \\
100 UCLA Medical Plaza, Suite 550, Los Angeles, CA 90095--7059, USA \\
Tel.: +1/310/825--9203 \\
Fax: +1/310/443--0477}

\vspace{0.5in}

% contact information for authors
\noindent {\scriptsize E-mail addresses:
\href{mailto:alanderos@ucla.edu}{alanderos@ucla.edu} (A. Landeros),
\href{mailto:stutztim@ucla.edu}{stutztim@ucla.edu} (T. Stutz),
\href{mailto:klkeys@g.ucla.edu}{klkeys@g.ucla.edu} (K. L. Keys),
\href{mailto:alekseye@musc.edu}{alekseye@musc.edu} (A. Alekseyenko),
\href{mailto:JanetS@mednet.ucla.edu}{JanetS@mednet.ucla.edu} (J. S. Sinsheimer),
\href{mailto:klange@ucla.edu}{klange@ucla.edu} (K. Lange),
\href{mailto:msehl@mednet.ucla.edu}{msehl@mednet.ucla.edu} (M. E. Sehl)}

\clearpage

%%%%% body of text %%%%%
\section*{Abstract}

\noindent \textbf{Background and Objectives}:
Biological systems with intertwined feedback loops pose a challenge to mathematical modeling efforts.
Moreover, rare events, such as mutation and extinction, complicate system dynamics.
Stochastic simulation algorithms are useful in generating time-evolution trajectories for these systems because they can adequately capture the influence of random fluctuations and quantify rare events.
We present a simple and flexible package, \pkg{BioSimulator.jl}, for implementing the Gillespie algorithm, \( \tau \)-leaping, and related stochastic simulation algorithms.
The objective of this work is to provide scientists across domains with fast, user-friendly simulation tools.

\noindent \textbf{Methods}:
We used the high-performance programming language \proglang{Julia} because of its emphasis on scientific computing.
Our software package implements a suite of stochastic simulation algorithms based on Markov chain theory.
We provide the ability to (a) diagram Petri Nets describing interactions, (b) plot average trajectories and attached standard deviations of each participating species over time, and (c) generate frequency distributions of each species at a specified time.

\noindent \textbf{Results}:
\pkg{BioSimulator.jl}'s interface allows users to build models programmatically within \proglang{Julia}.
A model is then passed to the \code{simulate} routine to generate simulation data.
The built-in tools allow one to visualize results and compute summary statistics.
Our examples highlight the broad applicability of our software to systems of varying complexity from ecology, systems biology, chemistry, and genetics.

\noindent \textbf{Conclusion}:
The user-friendly nature of \pkg{BioSimulator.jl} encourages the use of stochastic simulation, minimizes tedious programming efforts, and reduces errors during model specification.

\noindent \textbf{Keywords}: stochastic simulation, Gillespie algorithm, $\tau$-leaping, systems biology, Julia language

\section{Introduction}
Biological systems with overlapping feedback and feedforward loops are often inherently stochastic.
Furthermore, large, complex systems are mathematically intractable, and dynamical predictions based on deterministic models can be grossly misleading \citep{adams-arkin,elowitz2002}.
Stochastic simulation algorithms based on continuous-time Markov chains allow researchers to generate accurate time-evolution trajectories, test the sensitivity of models to key parameters, and quantify frequencies of rare events \citep{gillespie76,wilkinson06,elsamad,voliotis2016}.
Stochastic simulation is helpful in cases where (a) rare events, such as extinction or mutation, influence system dynamics, (b) population compartments, such as numbers of biochemical molecules, are present in small numbers, and (c) population cycles arise from demographic stochasticity.
Examples of such systems include gene expression networks, tumor suppressor pathways, and demographic and ecological systems.
The current paper introduces a simple and flexible package, \pkg{BioSimulator.jl}, for implementing popular stochastic simulation algorithms based on Markov chain theory.
\pkg{BioSimulator.jl} builds on previous software.
Notable examples include:
\begin{itemize}
  \item \pkg{StochSS}, an integrated framework for deploying simulations on high-performance clusters \citep{stochss}.
  It features a graphical interface for model editing, tools for deterministic, stochastic, and spatial simulations, a model analysis toolkit, and data visualization.
  \pkg{StochKit2}, a mature \proglang{C++} library of stochastic simulation algorithms, serves as the simulation engine.

  \item \pkg{StochPy}, an interactive stochastic modeling tool written in \proglang{Python} \citep{stochpy}.
  It provides a number of simulation algorithms, including delayed and single molecule methods.

  \item \pkg{GillespieSSA}, an \proglang{R} implementation of exact and approximate simulation algorithms \citep{gillespieSSA}.
  \pkg{Gillespie.jl} is a \proglang{Julia} extension of the original \proglang{R} package \citep{gillespiejl}.
  Today, \pkg{Gillespie.jl} features Jensen's uniformization method and the True Jump Method.

  \item \pkg{DifferentialEquations.jl}, an extensive \proglang{Julia} ecosystem for solving differential equations \citep{diffeq2017}.
\end{itemize}
In addition, there are many specialized biological modeling tools:
\begin{itemize}
  \item \pkg{COPASI}, a large open-source software application for analyzing and simulating biochemical networks models \citep{copasi}. It features a user-friendly graphical interface and supports both differential equation modeling and stochastic simulation algorithms.

  \item \pkg{Smoldyn}, a particle-based stochastic spatial simulation engine \citep{smoldyn}.
  It emphasizes biophysical and cell environment modeling.

  \item \pkg{BioNetGen}, a rules-based model editor and simulation framework that focuses on cell regulatory networks \citep{harris2016}.

  \item \pkg{PySB}, a \proglang{Python} module for building mathematical descriptions of biological networks \citep{lopez2013}.
  It provides a domain-specific language that is translated into a portable intermediate model representation.
  \pkg{PySB} connects to other software tools, such as \pkg{SciPy} and \pkg{StochKit2}, for simulations.
\end{itemize}
These packages have grown more sophisticated over time.
Our goal in developing \pkg{BioSimulator.jl} is to provide a fast, open-source, and user-friendly library of stochastic simulation algorithms.

% why are these algorithms popular?
 \pkg{BioSimulator.jl} is written in the high-level, high-performance programming language \proglang{Julia} \citep{bezanson12}.
 Our software consists of three main components: an interactive interface for model prototyping, a simulation engine, and a small library of stochastic simulation algorithms.
 We briefly review the theory underlying stochastic simulation in Section 2 and present the algorithms implemented by \pkg{BioSimulator.jl} in Section 3.
 Section 4 outlines the model development process and describes the available visualization tools.
 We summarize \pkg{BioSimulator.jl}'s graphical inputs and outputs, including (a) Petri nets describing the connectivity of the reaction network, (b) time-evolution trajectories of the system, and (c) frequency distributions of events.
 Three examples from biology, chemistry, and genetics in Section 5 illustrate how to define a model in \pkg{BioSimulator.jl} and simulate it as a continuous-time Markov chain.
 \pkg{BioSimulator.jl} will be valuable to a broad range of molecular and systems biologists, physicists, chemists, applied mathematicians, statisticians, and computer scientists interested in stochastic simulation modeling.

\section{Background}

\subsection{Markov jump processes}

Before we discuss simulation specifics, we describe the time evolution of a Markov jump process \citep{Lange2003}.
The underlying Markov chain follows a column vector $\mathbf{X}_{t}$ whose $i$-th component ${X}_{ti}$ is the number of particles of type $i$ at time $t \ge 0$.
The components of $\mathbf{X}_{t}$ track species counts and are necessarily non-negative integers.
The system starts at time $0$ and evolves via a succession of random reactions.
Let $c$ denote the number of reaction channels and $d$ the number of particle types.
Channel $j$ is characterized by a propensity function $r_{j}(\mathbf{x})$ depending on the current vector of counts $\mathbf{x}$. 
In a small time interval of length $s$, we expect $r_{j}(\mathbf{x})s + o(s)$ reactions of type $j$ to occur.
Reaction $j$ changes the count vector by a fixed integer vector $\mathbf{v}^{j}$.
Some components $v_{k}^{j}$ of $\mathbf{v}^{j}$ may be positive, some 0, and some negative.

From the wait and jump perspective of Markov chain theory, the chain waits an exponential length of time until the next reaction. 
If the chain is currently in state $\mathbf{x} \equiv \mathbf{X}_{t}$, then the intensity of the waiting time until the next reaction is $r_{0}(\mathbf{x}) = \sum_{j=1}^{c} r_{j}(\mathbf{x})$.
Once the decision to jump is made, the chain jumps to the neighboring state $\mathbf{x} + \mathbf{v}^{j}$ with probability $r_{j}(\mathbf{x}) / r_{0}(\mathbf{x})$.
Table~\ref{tab:reactions} lists typical reactions, their propensities $r(\mathbf{x})$, and increment vectors $\mathbf{v}$.
In the table, $S_{i}$ denotes a single particle of type $i$.
Only the nonzero increments $v_{i}$ are shown.
The reaction propensities invoke the law of mass action and depend on rate constants $a_{i}$ \citep{highham}.
Each discipline has its own vocabulary.
Chemists use the term propensity instead of the term intensity and call the increment vector a stoichiometric vector.
Physicists prefer creation to immigration.
Biologists speak of death and mutation rather than of decay and isomerization.
\begin{table}
\centering
\ra{1.3}
\begin{tabular}{@{}rlll@{}}
    \toprule
    Name & Reaction & $r(\mathbf{x})$ & $\mathbf{v}$ \\
    \midrule
    Immigration & $0 \to S_{1}$ & $a_{1}$ & $v_{1} = 1$ \\
    Decay & $S_{1} \to 0$ & $a_{2} x_{1}$ & $v_{1} = -1$ \\
    Dimerization & $S_{1} + S_{1} \to S_{2}$ & $a_{3}\binom{x_{1}}{2}$ & $v_{1} = -2, v_{2} = 1$ \\
    Isomerization & $S_{1} \to S_{2}$ & $a_{4}x_{1}$ & $v_{1}=-1, v_{2}=1$ \\
    Dissociation & $S_{2} \to S_{1} + S_{1}$ & $a_{5}x_{2}$ & $v_{1}=2, v_{2}=-1$ \\
    Budding & $S_{1} \to S_{1} + S_{2}$ & $a_{6}x_{1}$ & $v_{2}=1$ \\
    Replacement & $S_{1} + S_{2} \to S_{2} + S_{2}$ & $a_{7}x_{1}x_{2}$ & $v_{1}=-1, v_{2}=1$ \\
    Complex Reaction & $S_{1} + S_{2} \to S_{3} + S_{4}$ & $a_{8}x_{1}x_{2}$ & $v_{1}=v_{2}=-1, v_{3}=v_{4}=1$\\
    \bottomrule
\end{tabular}
\caption{Propensities $r(\mathbf{x})$ and increment vectors $\mathbf{v}$ for some typical reactions.
Here, $S_{i}$ denotes a single particle of type $i$, and $a_{i}$ denotes the reaction rate constant.\label{tab:reactions}}
\end{table}

% Need to provide background for the chemical master equation here
Often one is interested in the finite-time transition probabilities of a Markov chain.
Let $p_{\mathbf{x},\mathbf{y}}(t)$ denote the probability that a Markov chain starting at state $\mathbf{x}$ transitions to state $\mathbf{y}$ by time $t$.
The \textit{chemical master equation} reads
\[
    p_{\mathbf{x},\mathbf{y}}(t + dt) =  p_{\mathbf{x},\mathbf{y}}(t)\left[1 - \sum_{j=1}^{c} a_{j}(\mathbf{y}) dt\right] + \sum_{j=1}^{c} p_{\mathbf{x},\mathbf{y} - \mathbf{v}^{j}}(t)\ a_{j}(\mathbf{y} - \mathbf{v}^{j}) dt + o(dt),
\]
or equivalently, in differential form
\[
    \frac{d}{dt} p_{\mathbf{x},\mathbf{y}}(t) = \sum_{j=1}^{c}\left[ p_{\mathbf{x},\mathbf{y} - \mathbf{v}^{j}}(t)\ a_{j}(\mathbf{y} - \mathbf{v}^{j}) - p_{\mathbf{x},\mathbf{y}}(t)\ a_{j}(\mathbf{y}) \right].
\]
The differential form is a system of possibly infinitely-many coupled differential equations.
Solving the master equation is an enormous endeavor except in the simplest of models.
Alternatively, simulating multiple realizations of the process yields data that can then be used to estimate transition probabilities and summary statistics.
References~\citep{gillespie76,gillespie77} offer a first-principles derivation of the chemical master equation for biochemical reactions and connect this formalism to simulation methods.

\subsection{The Julia language}

% make sure we cite the SIAM paper and the more technical arXiv paper
\proglang{Julia} is a fast, expressive, and flexible programming language for scientific computing \cite{bezanson12}.
Specifically, the language targets the so-called \textit{two-language problem}, in which a methods developer builds working prototypes in a \textit{slow} high-level language only to then move performance-critical subroutines to a \textit{fast} low-level language.
The implicit assumption is that high-level, dynamic languages are expressive and therefore easier to use but at the expense of performance.
For example, some language features, such as ``for loops'', may incur performance penalties through no fault of the user.
Such performance hurdles are overlooked because dynamic programming languages typically avoid tedious compilation steps and provide users with an intuitive syntax.
High-level languages allow users to express the tasks they wish to complete by handling the low-level details.
On the other hand, low-level languages typically require technical expertise and offer the fastest possible execution.

A consequence of the two-language problem is that many tools, especially scientific software, become fragmented and difficult to manage.
This is all the more relevant to scientists who often lack the necessary skills to maintain a software engineering project.
\proglang{Julia} tackles this problem by providing an intuitive syntax and language features compatible with high performance underneath the hood, thereby making the user all the more productive.
Interested readers are encouraged to peruse \citep{bezanson12} for a deeper look into the \proglang{Julia} philosophy.
Many online tutorials for learning \proglang{Julia} are available at \url{https://julialang.org/learning/}.

\section{Simulation methods}

\pkg{BioSimulator.jl} supports five different simulation algorithms.
In the following subsections, we review each algorithm and briefly describe its strengths and weaknesses.
The purpose of this section is to provide the reader with a high-level understanding of each algorithm and develop an intuition as to where methods succeed and fail.
There are many references that provide details on these simulation methods, elucidate connections to spatial systems and related stochastic models, and motivate applications in many fields~\citep{stoch-review, Golightly2013}.

\subsection{Stochastic simulation algorithm}

The Stochastic Simulation Algorithm (SSA), also known as the Direct or Gillespie method, implements the wait and jump mechanism for simulating a continuous-time Markov chain \citep{gillespie76,gillespie77}.
At each step, the algorithm computes the propensities $r_{j}(\mathbf{x})$ of each reaction channel and generates two random deviates.
One of these is an exponential deviate indicating the time to the next reaction based on $r_{0}(\mathbf{x})$.
The second is a uniform random number $U(0,1)$ determining which reaction fires next based on the ratios $r_{j}(\mathbf{x})/r_{0}(\mathbf{x})$.
The two main computational steps in the SSA are
\begin{enumerate}[label=(\Roman*)]
    \item Generate a \textit{random time} $s$ to the next event by sampling from an exponential distribution with rate $r_{0}(\mathbf{x})$.
    This provides the update $t \mapsto t + s$.

    \item Generate a \textit{random index} $j$ denoting the reaction that occurred by sampling a categorical distribution with probabilities $r_{j}(\mathbf{x}) / r_{0}(\mathbf{x})$.
    This provides the update $\mathbf{x} \mapsto \mathbf{x} + \mathbf{v}^{j}$.
\end{enumerate}
Since the propensities $r_{j}(\mathbf{x})$ change after each event, the distributions underlying steps (I) and (II) change over time.
Every Gillespie-like simulation algorithm is effectively distinguished by the sampling procedures used to generate the required random numbers, and the algorithms that update the underlying probability distributions.

The main advantage of the SSA is its ability to produce statistically correct trajectories and distributions by simulating every reaction.
This strength is also its greatest weakness in models where a small subset of frequently occurring reactions dominate simulation.
The detailed computational analysis of \citeauthor{odmpaper} identifies the linear search on the propensities $r_{j}(\mathbf{x})$ as a major obstacle to fast simulation \citep{odmpaper}.
The algorithm does not scale well with model size in the presence of different time scales.
Thus, one must balance the value of accurate results versus speed in selecting SSA for simulation.

% We need a citation for FRM!
\subsection{First reaction method}
Gillespie proposed the First Reaction Method (FRM) as an alternative to SSA\@ \citep{gillespie76}.
The main difference is the time to the next reaction
\[ \tau = \underset{1\le j \le c}{\min}\{w_{1}, \ldots, w_{c}\} \]
defined by independent exponentially distributed waiting times $w_{1}, \ldots, w_{c}$ with intensity $r_{j}$.
Here $c$ again denotes the total number of reaction channels.
The premise of the algorithm is to compute the minimum of $c$ exponential random variables explicitly.
This approach is less computationally efficient than SSA in the number of exponential deviates required to compute the time to the next event.
We include the FRM in \pkg{BioSimulator.jl} purely for educational purposes.
While the FRM does not offer any advantage over the original SSA, it provides a different way of thinking about simulation.
The Next Reaction Method builds upon this idea.

\subsection{Next reaction method} The Next Reaction Method (NRM), also known as the Gibson-Bruck method, is another exact algorithm equivalent to SSA \citep{gibsonbruck}.
At time $t = 0$, the algorithm seeds each reaction channel $j$ a firing time $\tau_{j}$ and stores them inside a \textit{priority queue}.
In this context, a priority queue is a data structure that sorts pairs $(j, \tau_{j})$ according to the value of $\tau_{j}$ in increasing order.
That is, if $\tau_{J}$ is the minimum time, then the pair $(J, \tau_{J})$ appears at the top of the queue.
Thus, the next reaction is $J$ and its firing time is $\tau_{J}$; all other reactions fire at some future time.
After reaction $J$ fires, the NRM updates the state vector $\mathbf{x} \to \mathbf{x} + \mathbf{v}^{J}$.
The next firing time $\tau_{J}$ is also updated by an appropriate exponential deviate:
\[
    \tau_{J,\mathrm{new}} = \tau_{J,\mathrm{old}} + \mathrm{Exponential}(r_{J,\mathrm{new}}(\mathbf{x})),
\]
where $r_{J,\mathrm{new}}(\mathbf{x})$ is the new propensity value.
The remaining firing times change according to the recipe
\[ \tau_{j,\mathrm{new}} = t + \frac{r_{j,\mathrm{old}}}{r_{j,\mathrm{new}}}(\tau_{j,\mathrm{old}} - t), \quad j \neq J,
\]
based on the lack of memory property of the exponential distribution.

The NRM also minimizes the number of propensities updated by tracking dependencies between reaction channels.
Typically, the SSA sweeps through all the  propensities $r_{j}(\mathbf{x})$ to reflect the change in $\mathbf{x}$.
However, a reaction channel's propensity only changes when the previous reaction event affected components of $\mathbf{x}$ that appears as reactants.
The NRM uses a \textit{reaction dependency graph} to describe these relationships between reactions.
This data structure reduces the number of propensities that must be updated.
Kahan summation can also be used to update the cumulative intensity $r_{0}(\mathbf{x})$ efficiently \citep{gibsonbruck,mauch2011}.

The NRM excels in simulating systems with large numbers of species and lightly coupled reactions.
Otherwise, in the extreme case, the algorithm must recalculate every firing time at every step.
Systems with heavily coupled reaction channels are problematic for the NRM \citep{odmpaper}.
In this setting, the NRM becomes identical to the SSA but with the added burden of maintaining its priority queue.

\subsection{Optimized direct method}
The Optimized Direct Method (ODM) improves upon the original SSA by exploiting multi-scale properties inherent in large models \citep{odmpaper}.
\pkg{BioSimulator.jl}'s ODM implementation simulates a system once to count the number of times each reaction fires.
This allows one to classify each reaction channel as fast (high frequency) or slow (low frequency).
Sorting the reactions from fast to slow reduces the search depth in selecting the next reaction.
This approach works well with heavily coupled reactions.
Some systems exhibit more erratic behavior that prohibits classifying a reaction fast or slow.
That is, switching between different time scales thwarts the ODM's sorting optimization.
The auto-regulation genetic network in Example 3 is an example of a system that undermines the optimized sorting of ODM\@.

\subsection[Tau-leaping]{$\tau$-leaping}
\pkg{BioSimulator.jl} implements performance optimizations described by \citeauthor{mauch2011} to improve SSA, FRM, NRM, and ODM techniques \citep{mauch2011}.
However, algorithms that simulate every reaction ultimately succumb to the high computational expense of large models.
The $\tau$-leaping framework attempts to accelerate simulation by lumping reaction events together within a time leap $\tau$, selected to be as large as possible \citep{gillespie01,gillespie03,efficient}.
The basic $\tau$-leaping formula is
\[
    \mathbf{X}_{t+\tau} = \mathbf{X}_{t} + \sum_{j=1}^{c} \mathbf{v}^{j} Y_{j}(r_{j}(\mathbf{X}_{t}) \tau).
\]
where $Y_{j}$ is a Poisson random variable with rate $r_{j}(\mathbf{X}_{t}) \tau$.
Thus, $\tau$-leaping accelerates the SSA by lumping together multiple reaction events over an interval of size $\tau$.
The main challenge in $\tau$-leaping is selecting the step size as large as possible while satisfying the \textit{leap condition}
\[
    \abs{r_{j}(\mathbf{X}_{t+\tau}) - r_{j}(\mathbf{X}_{t})} \le \epsilon, \quad j = 1,2,\ldots,c,
\]
which states that the propensity for each reaction $j$ is approximately constant over a leap of size $\tau$.
Here, $\epsilon \in (0,1]$ is a prescribed acceptable change in propensities that \textit{controls} the accuracy of sample paths generated by a $\tau$-leaping algorithm.
A larger $\epsilon$ allows for larger leaps, while a smaller $\epsilon$ restricts leap size.
In practice, many $\tau$-leaping algorithms employ a surrogate condition that satisfies the leap condition with high probability.

In the stochastic simulation setting, a system is said to be \textit{stiff} if the dynamics force a simulator to take ``small'' steps.
Stiffness arises for a variety of reasons. % citations!!!
Large models typically have a number of reactions occurring within a given interval.
Reactions occurring on separate time scales split the system between ``fast'' and ``slow'' reactions, with the former occurring in a nearly deterministic fashion.
In any case, stiffness causes the number of simulated events to increase in exact methods like the SSA.
Stiffness poses a second threat to $\tau$-leaping methods.
In addition to decreasing the leap size, stiffness can cause $\tau$-leaping to generate an excess of events due to the unbounded nature of the Poisson distribution.

There are two precautionary measures to protect against aberrant behavior in $\tau$ selection \citep{avoiding,sehl09b}.
For example, a $\delta$ parameter controls whether $\tau$-leaping will default to SSA when the leap size is less than the expected change under the SSA; that is, if $\tau < \delta \cdot 1 / r_{0}(\mathbf{x})$.
This precaution is necessary to avoid taking suboptimal steps that introduce error and to mitigate leaps that send the system into negative population counts.
In the event of a negative excursion, an acceptance parameter $\beta$ in $(0,1)$ contracts the leap step, effectively thinning the number of reaction events.
Specifically, each event in a bad leap is randomly accepted if a uniform deviate $U(0,1)$ is less than $\beta$.
The leap size $\tau$ is then set to $\tau \mapsto \beta \tau$.
Leap contraction introduces bias in sample paths, so one must take care in setting $\beta$.
As a rule of thumb, one should first select conservative values for $\epsilon$, $\beta$, and $\delta$ and test performance using short numerical experiments.
\pkg{BioSimulator.jl} sets $\epsilon = 0.03$, $\beta = 0.75$, and $\delta = 2$ as default values, drawn from the literature, that ought to perform well in many cases.

$\tau$-leaping discards reaction event times but reduces the burden of random number generation.
Each leap in the algorithm requires $c$ Poisson random deviates, one for each reaction channel.
This accelerates simulation when the leap size $\tau$ is significantly large compared to a single SSA step.
Like the SSA, there are many sophisticated variations on the original $\tau$-leaping algorithm.
\pkg{BioSimulator.jl} implements the version found in~\citep{gillespie03} and is referred to as Ordinary $\tau$-Leaping (OTL).
Future development will implement additional $\tau$-leaping algorithms from the literature.
The next section reviews Step anticipation $\tau$-leaping, a second $\tau$-leaping method.

\subsection[Step anticipation tau-leaping]{Step anticipation $\tau$-leaping}

The Step Anticipation $\tau$-Leaping (SAL) algorithm is a variation on $\tau$-leaping \citep{sehl09b}.
In the SAL algorithm, one approximates each propensity by a first-order Taylor polynomial around $t$ with starting value $r_j(\mathbf{x})$.
The number of reactions of type $j$ is then sampled from a Poisson distribution with mean
\begin{equation*}
\omega_j(t,t+\tau) = \int_0^t \left[ r_j(\mathbf{x}) + \frac{d}{dt}r_j(\mathbf{x})s\right] \mathrm{d}s
= r_j(\mathbf{X})\tau + \frac{d}{dt}r_j(\mathbf{x})\frac{1}{2} \tau^2.
\end{equation*}
The deterministic reaction rate equation
\begin{eqnarray*}
    \frac{d}{dt}x_{k} &=& \sum_{j=1}^{c} r_{j}(\mathbf{x}) v_{j}^{k} \\
\end{eqnarray*}
allows one to approximate the derivatives $\frac{d}{dt}r_j(\mathbf{x})$ at $\mathbf{x}$ by applying the chain rule of differentiation:
\begin{eqnarray*}
\frac{d}{dt}r_j(\mathbf{x}) = \sum_{k=1}^{d} \frac{\partial}{\partial x_{k}} r_{j}(\mathbf{x})\frac{d}{dt} x_{k} \approx \sum_{k=1}^{d} \frac{\partial}{\partial x_{k}} r_{j}(\mathbf{x})\sum_{i=1}^{c} r_{i}(\mathbf{x})v_{i}^{k}.
\end{eqnarray*}
The required partial derivatives $\frac{\partial}{\partial x_{k}} r_{j}(\mathbf{x})$ are typically constant or linear for mass-action kinetics.
The main advantages of SAL are that it improves accuracy over other $\tau$-leaping algorithms without compromising speed.
This is crucial for complex systems exhibiting rapid fluctuations in reaction propensity and higher order kinetics.
The critical step in SAL is selecting the leap size $\tau$ so that the linear approximation to the system holds and avoids negative populations.
In our implementation, we select $\tau$ so that the bound
\[ \abs{\frac{d}{dt}r_{j}(\mathbf{x})} \tau \le \epsilon \max\{r_{j}(\mathbf{x}),c_{j}\} \]
holds for every reaction~\citep{gillespie03}.
Here $c_{j}$ is the rate constant of reaction $j$ and $\epsilon$ is the tuning parameter.
Our implementation of SAL includes the negative population safeguards outlined in the previous section.

\section{Software description}

This section briefly reviews \pkg{BioSimulator.jl}'s interface and its tools.
We refer interested readers to the package documentation for technical details.
One may access \pkg{BioSimulator.jl}'s documentation through \proglang{Julia}'s help system based on the convention \code{?<name>}, where \code{<name>} is the name of a function of interest, such as \code{sum}.

\subsection{Creating a model}
First, a user loads \pkg{BioSimulator.jl}'s interface, simulation routines, and other helper functions with the command \code{using BioSimulator}.
The \code{Network} construct is central to model specification.
It represents a system of interacting particles starting from some initial state $\mathbf{x}_{0}$.
A \code{Network} object stores the initial population sizes for each \code{Species} and the definitions for each \code{Reaction}.
One constructs a \code{Network} by passing a name to the system and successively adding each component with the \code{<=} symbol.
For example, the following code
\begin{minted}[fontsize=\scriptsize]{julia}
    model = Network("Michaelis-Menten")
    
    model <= Species("S", 301)
    model <= Species("E", 130)
    model <= Species("SE",  0)
    model <= Species("P",   0)
\end{minted}
defines four \code{Species} named \textit{S} (substrate), \textit{E} (enzyme), \textit{SE} (substrate-enzyme complex), and \textit{P} (protein) with initial counts $\mathbf{x}_{0} = \left(301, 130, 0, 0\right)$.
One defines a \code{Reaction} by providing a label, a reaction rate constant, and the reaction equation itself. For example, the code
\begin{minted}[fontsize=\scriptsize]{julia}
    model <= Reaction("dimerization", 0.00166, "S + E --> SE")
    model <= Reaction("dissociation", 0.0001,  "SE --> S + E")
    model <= Reaction("conversion",   0.1,     "SE --> P + E")
\end{minted}
defines the dimerization, dissociation, and conversion reactions with rate constants $0.00166$, $0.0001$, and $0.1$, respectively.
\pkg{BioSimulator.jl} assumes mass action kinetics in simulations (see Table~\ref{tab:reactions}).

\subsection{Running simulations}
The \code{simulate} method parses a \code{Network} and carries out a simulation run using one of \pkg{BioSimulator.jl}'s algorithms.
The command to simulate is written as
\begin{minted}[fontsize=\scriptsize]{julia}
    simulate(model, algname, output_type = Val(:fixed), time = 1.0, epochs = 1, trials = 1, kwargs...).
\end{minted}
Here \code{model} and \code{algname} are the required inputs.
The remaining inputs are \textit{keyword arguments} invoked by key-value pairs; each of these optional arguments has a default value.
We summarize these inputs below along with any default values:
\begin{itemize}
    \item \code{model}: The \code{Network} to simulate.
    
    \item \code{algname}: A simulation algorithm.
    One may choose between \code{Direct()}, \code{FirstReaction()},
    \code{NextReaction()}, \code{OptimizedDirect()}, \code{TauLeaping()}, or \code{StepAnticipation()}.
    
    \item \code{output_type = Val(:fixed)}: One of \code{Val(:fixed)} or \code{Val(:full)} denoting a strategy for saving the state vector.
    The \code{Val(:full)} option has the simulator sample the state vector after each reaction event and records its value.
    This option uses more memory and may incur a slight performance penalty due to the fact that \pkg{BioSimulator.jl} cannot determine \textit{in advance} the size of the output.
    The \code{Val(:fixed)} option records the state vector at fixed intervals.
    
    \item \code{time = 1.0}: The amount of time to simulate the model.
    If one specifies a time $t$, then the model will be simulated over the interval $(0,t)$.
    
    \item \code{epochs = 1}: The number of save points when using fixed-interval output (\code{Val(:fixed)}).
    The default option \code{epochs = 1} records the initial and final values of the state vector.
    
    \item \code{trials = 1}: The number of independent realizations to simulate.

    \item \code{track_stats = false}: A Boolean value that indicates whether the simulation should keep track of algorithm-specific statistics.
    Using the option with SSA-like methods simply track the number of events.
    $\tau$-leaping methods also include the number of times the simulation encounters negative population counts.
    
    \item \code{kwargs}: A catch-all for additional options specific to an algorithm.
    One can check these options using \code{?<algname>}, where \code{<algname>} is to be replaced with an algorithm type.
    For example, \code{?TauLeaping} will print a description of the simulation method.
\end{itemize}
As an example, the code
\begin{minted}[fontsize=\scriptsize]{julia}
    simulate(model, StepAnticipation(), Val(:fixed), time=100.0, epochs=50, trials=1000, epsilon = 0.125)
\end{minted}
will simulate the given model with SAL and return fixed-interval output.
The time interval $(0, 100)$ is discretized into $50$ epochs for each of the $1000$ independent realizations of the stochastic process.
Lastly, the \code{epsilon = 0.125} option specifies the value of the $\epsilon$ parameter used by SAL to control leap size.

\subsection{A note on epochs}
By default, \pkg{BioSimulator.jl} partitions the simulation time span into \textit{epochs} of equal length.
After each simulation step, \pkg{BioSimulator.jl} checks whether the previous event pushed the simulation into the next epoch.
If so, it will record the current value of $\mathbf{x}$ at each of the previous epochs.
We note that the \code{Val(:fixed)} option \textit{does not} affect how each algorithm steps through a simulation.
However, this strategy necessarily discards information, such as waiting times between reactions.
Users must also take care to use a sufficiently large number of epochs so that the simulation data accurately captures system dynamics.
In particular, one may fail to capture phenomena occurring on a time scale smaller than the one implied by the number of epochs.
Despite its drawbacks, \pkg{BioSimulator.jl} favors fixed interval sampling because it assists in computing summary statistics, improves performance for long simulation runs, and facilitates interactive model prototyping.
The developers of \pkg{StochPy} provide an excellent discussion on the trade-offs between fixed-interval and full simulation output in \citep{stochpy}.

\subsection{Running parallel simulations}
Simulating large numbers of realizations is naturally amenable to parallelization because trial runs are independent.
\pkg{BioSimulator.jl} takes advantage of \proglang{Julia}'s built-in parallelism to speed up large simulation tasks.
This is achieved by specifying \code{julia --procs=N} when starting \proglang{Julia}.
Here $N$ is the number of worker threads.
Running \proglang{Julia} in parallel mode allows \pkg{BioSimulator.jl} to simulate a \code{Network} by delegating work to separate processes.
For example, if one has specified $4$ threads then \pkg{BioSimulator.jl} will simulate $1000$ realizations by delegating $250$ trials to each thread.
In practice, simulations of large networks benefit more from parallelization than small networks because generating a single trajectory is typically more expensive in the former scenario.

\subsection{Petri nets}
\begin{figure}[hp]
    \centering
    %%%%% Panel A %%%%%
    \begin{subfigure}[b]{0.2\textwidth}
        \includegraphics[width=\textwidth]{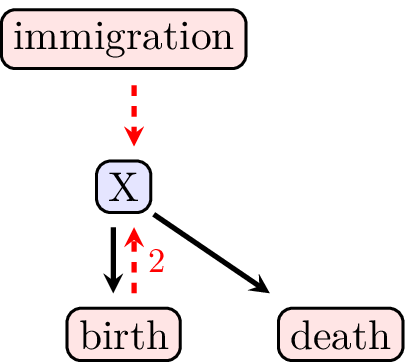}
        \caption{Kendall's process}
    \end{subfigure}
    \hspace{1em}
    %%%%% Panel B %%%%%
    \begin{subfigure}[b]{0.4\textwidth}
        \includegraphics[width=\textwidth]{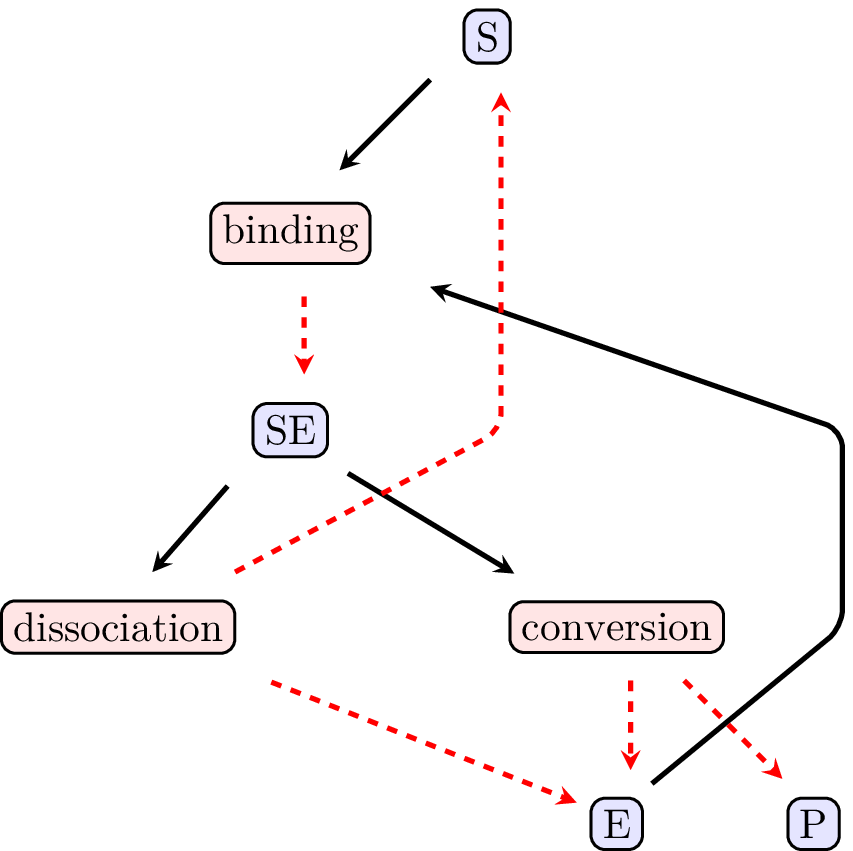}
        \caption{Michaelis-Menten enzyme kinetics}
    \end{subfigure}
    \vspace{1em} \\
    %%%%% Panel C %%%%%
    \begin{subfigure}[b]{0.9\textwidth}
        \includegraphics[width=\textwidth]{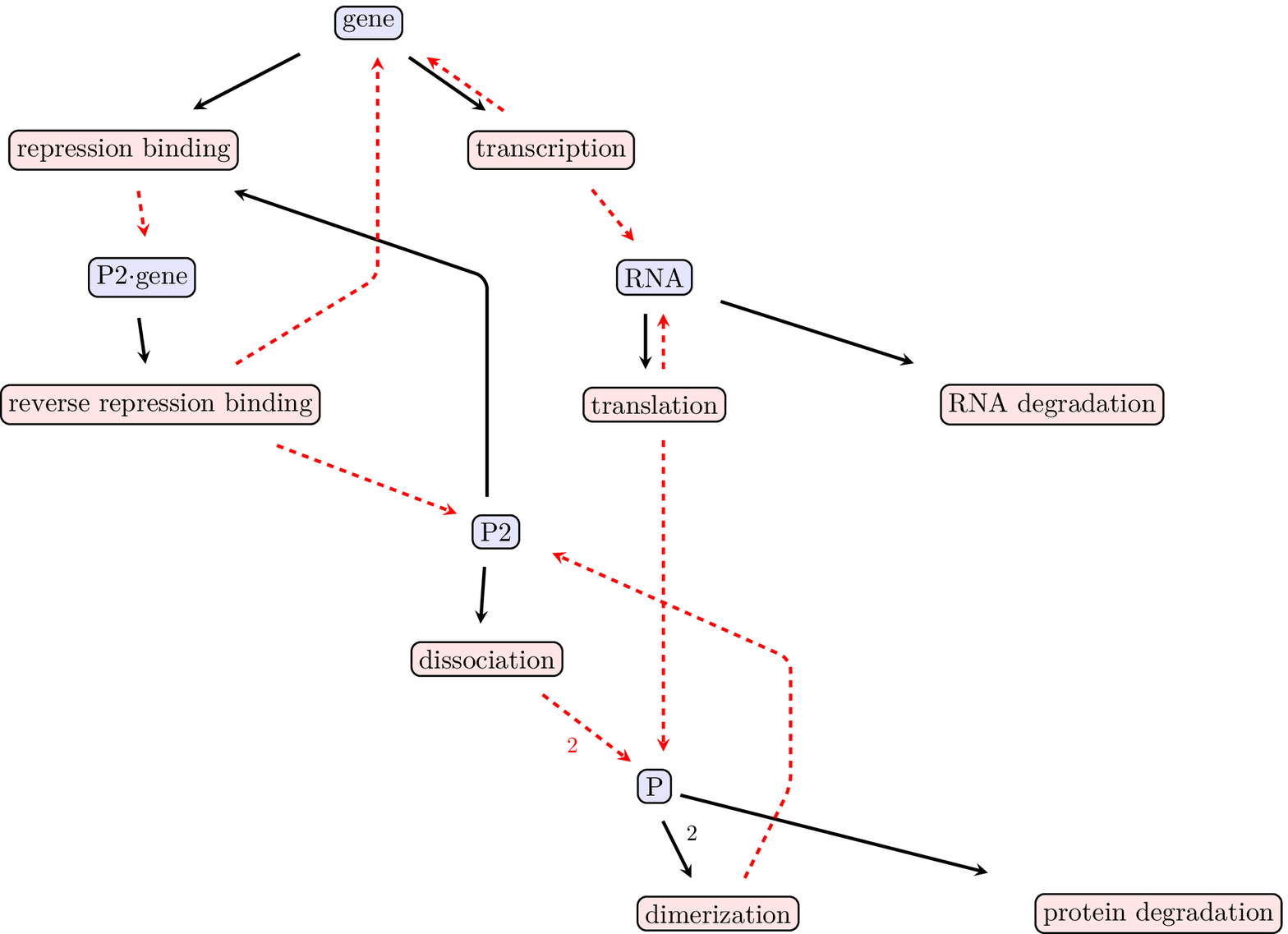}
        \caption{Prokaryotic autoregulation}
    \end{subfigure}
    \caption{Petri net representations of (a) Kendall's process, (b) the Michaelis-Menten model, and (c) a self-regulating gene network generated by \pkg{BioSimulator.jl} via the \pkg{TikzGraphs} package.
    Arrows connecting species to reaction denote how species enter into a reaction, and arrows from a reaction to a species denote how species are produced or deleted by the reaction.
    When the reaction produces more than 1 particle of a given species, its coefficient appears along the arrow connecting reaction to product species.\label{fig:petrinets}}
\end{figure}

\pkg{BioSimulator.jl} allows users to visualize the structure of their models as Petri nets using the \code{visualize} function (see Figure~\ref{fig:petrinets}).
A Petri net is a directed graph whose nodes represent either a species (oval) or reaction (rectangle).
An arrow from a species to a reaction indicates that the species acts as a reactant (black arrows), while the opposite direction indicates a product (red arrows).

\subsection{Simulation output}
Output from stochastic simulation at a series of time points can be plotted as mean trajectories over time or as full distributions.
\pkg{BioSimulator.jl} generates time series data for each species and stores it in a \code{SimulationSummary} object, which also tracks the model used in the simulation, simulation parameters, and key algorithm statistics.
\pkg{BioSimulator.jl} provides a few convenient functions for visualization and summary statistics through the \code{SimulationSummary} construct.
For example, one may access simulation data as a \code{DataFrame} provided by the \pkg{DataFrames.jl} package in \proglang{Julia}.
The \pkg{DataFrames.jl} documentation provides examples for carrying out common operations, including data manipulation, computing summary statistics, and saving data to a file.
\code{DataFrame} conversion is achieved by calling \code{DataFrame(result)}, where \code{result} is a \code{SimulationSummary}.
The resulting table has three types of columns.
The \code{time} and \code{trial} columns indicate the time point and trial number of a record.
The remaining columns are labelled according to the species or compartment name.
One must load the \pkg{DataFrames.jl} package before converting simulation output to a \code{DataFrame}:
\begin{minted}[fontsize=\scriptsize]{julia}
    > using DataFrames
    > result = simulate(model, time = 50.0, epochs = 3, trials = 3)
    > DataFrame(result)
    | Row | time | SE | S   | P   | E   | trial |
    |-----|------|----|-----|-----|-----|-------|
    | 1   | 0.0  | 0  | 301 | 0   | 130 | 1     |
    | 2   | 25.0 | 66 | 46  | 189 | 64  | 1     |
    | 3   | 50.0 | 12 | 0   | 289 | 118 | 1     |
    | 4   | 0.0  | 0  | 301 | 0   | 130 | 2     |
    | 5   | 25.0 | 63 | 38  | 200 | 67  | 2     |
    | 6   | 50.0 | 9  | 0   | 292 | 121 | 2     |
    | 7   | 0.0  | 0  | 301 | 0   | 130 | 3     |
    | 8   | 25.0 | 64 | 29  | 208 | 66  | 3     |
    | 9   | 50.0 | 17 | 0   | 284 | 113 | 3     |
\end{minted}

\subsection{Plotting}

\pkg{BioSimulator.jl} provides convenient methods for visualizing simulation results through the \pkg{Plots.jl} package.
Installing the \pkg{Plots.jl} package provides one with default recipes for plotting individual realizations, mean trajectories, and frequency histograms.
The \code{plot} function acts on a \code{SimulationSummary} to produce a figure depending on the value of \code{plot_type}: the possible choices are \code{:trajectory}, \code{:meantrajectory}, and \code{:histogram}.
Consider the following examples:
\begin{itemize}
    \item \code{plot(result, plot_type=:trajectory, trial=1)} will plot the sample paths for each species based on the results of the first trial.

    \item \code{plot(result, plot_type=:meantrajectory, species=["S", "E"], epochs = 100)} will plot the mean trajectories for the species $S$ and $E$ based on $100$ epochs.
    Error bars represent one standard deviation from the mean.

    \item \code{plot(result, plot_type=:histogram)} will plot the distribution of each species at the end of a simulation, based on the number of trials.
\end{itemize}
Plotting options can be mixed and matches based on the interface provided by \pkg{Plots.jl}.
Users may consult the documentation of \pkg{Plots.jl} for help in further customizing figures.

\section{Results}

To illustrate the workflow of \pkg{BioSimulator.jl}, we provide three simple numerical examples using the SAL method unless otherwise specified.
In each case, we set $\epsilon = 0.03$, $\delta = 2$, and $\beta = 0.75$.

\begin{example}Kendall's process\end{example}

Kendall's birth, death, and immigration process is a continuous-time Markov chain governed by a birth rate $a_{1}$ per particle, a death rate $a_{2}$ per particle, and an immigration rate $a_{3}$.
Let $X_{t}$ denote the count of a species $S$ at time $t$. The events (reactions)
\begin{equation*}
\left\{
	\begin{aligned}
		S & \rightarrow S + S &\text{Birth}\\
		S & \rightarrow 0     &\text{Death}\\
		0 & \rightarrow S     &\text{Immigration}
	\end{aligned}
\right.
\end{equation*}
occur with propensities $a_{1} \cdot S$, $a_{2} \cdot S$, and $a_{3}$, respectively.
The Petri net in Figure~\ref{fig:petrinets}~(a) shows the relationships between the particles and reactions.
It allows the modeler to visualize the flow of particles through the network and check whether the set of reactions specified accurately captures the biological pathways being studied.

The following code simulates this model in \pkg{BioSimulator.jl}:
\begin{minted}[fontsize=\scriptsize,samepage]{julia}
    using BioSimulator

    model = Network("Kendall's Process")

    model <= Species("S", 5)

    model <= Reaction("Birth",       2.0, "S --> S + S")
    model <= Reaction("Death",       1.0, "S --> 0")
    model <= Reaction("Immigration", 0.5, "0 --> S")

    result = simulate(model, StepAnticipation(), time=4.0, epochs=40, trials=100_000)
\end{minted}

\begin{figure}[hp]
	\centering
  %%%%% Panel A %%%%%
  \begin{subfigure}[b]{0.49\textwidth}
    \includegraphics[width=\textwidth]{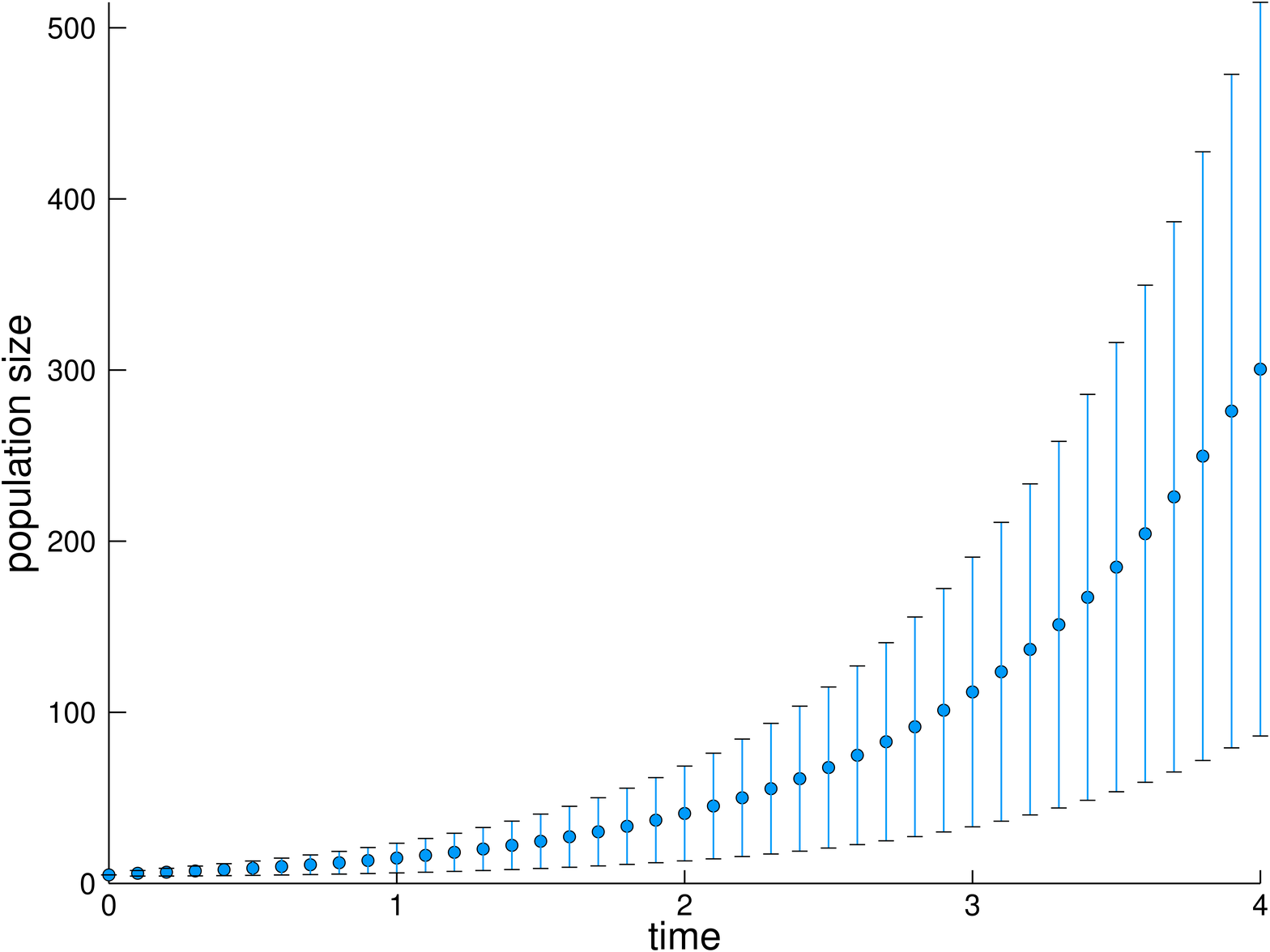}
    \caption{}
  \end{subfigure}
  %%%%% Panel B %%%%%
  \begin{subfigure}[b]{0.49\textwidth}
    \includegraphics[width=\textwidth]{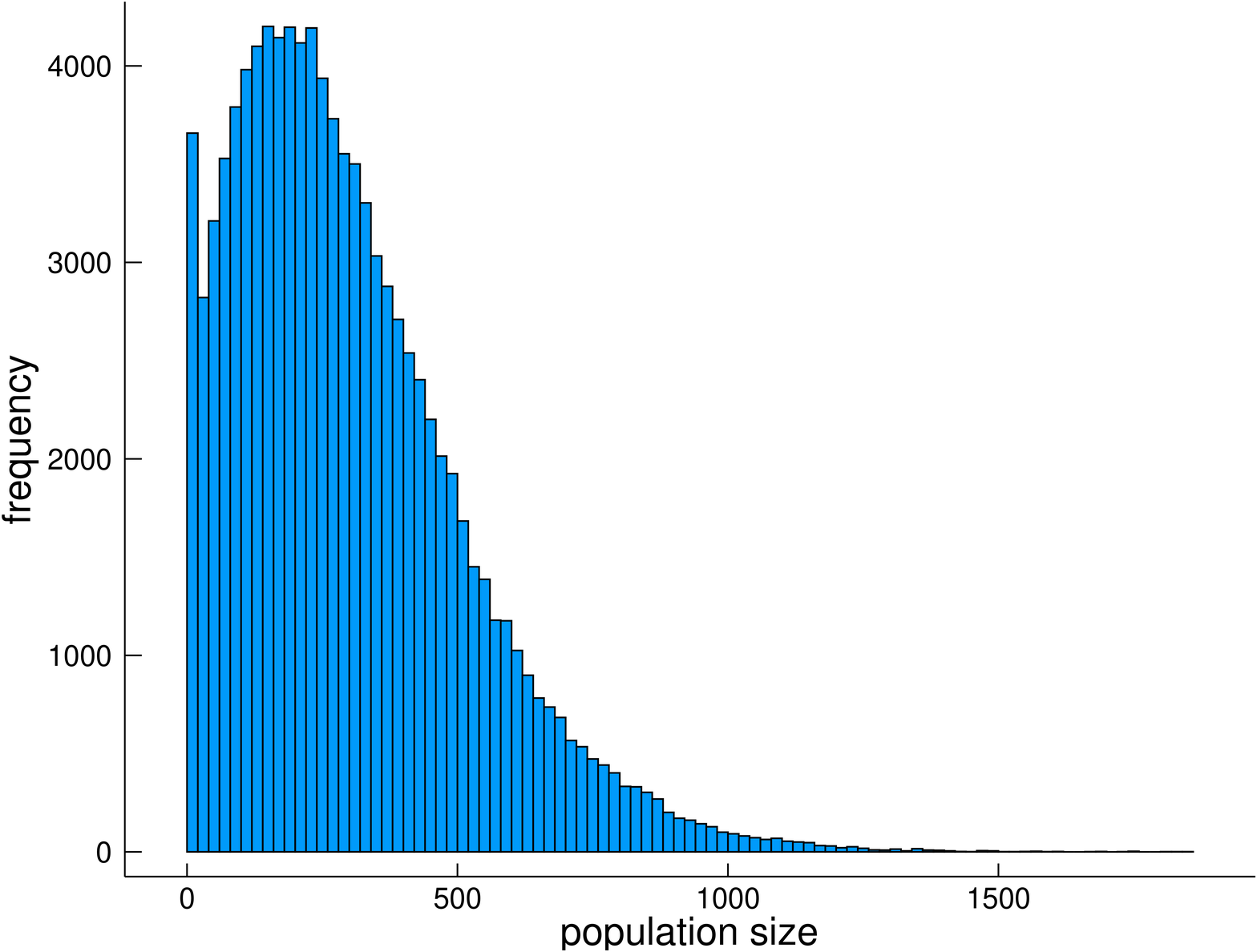}
    \caption{}
  \end{subfigure}
  \caption{Kendall's process with $\alpha = 2.0$, $\mu = 1.0$, and $\nu = 0.5$ starting from $5$ particles. (a) Mean trajectory and distribution of species $S$ computed from $10^5$ SAL realizations. The light, colored region represents one standard deviation from the mean at each recorded point. (b) The histogram suggests extinction is possible at $t=4$ even though the mean value is approximately $300$.\label{fig:kendall_plots}}
\end{figure}

\begin{figure}[hp]
	\centering
  %%%%% Panel A %%%%%
  \begin{subfigure}[b]{0.49\textwidth}
    \includegraphics[width=\textwidth]{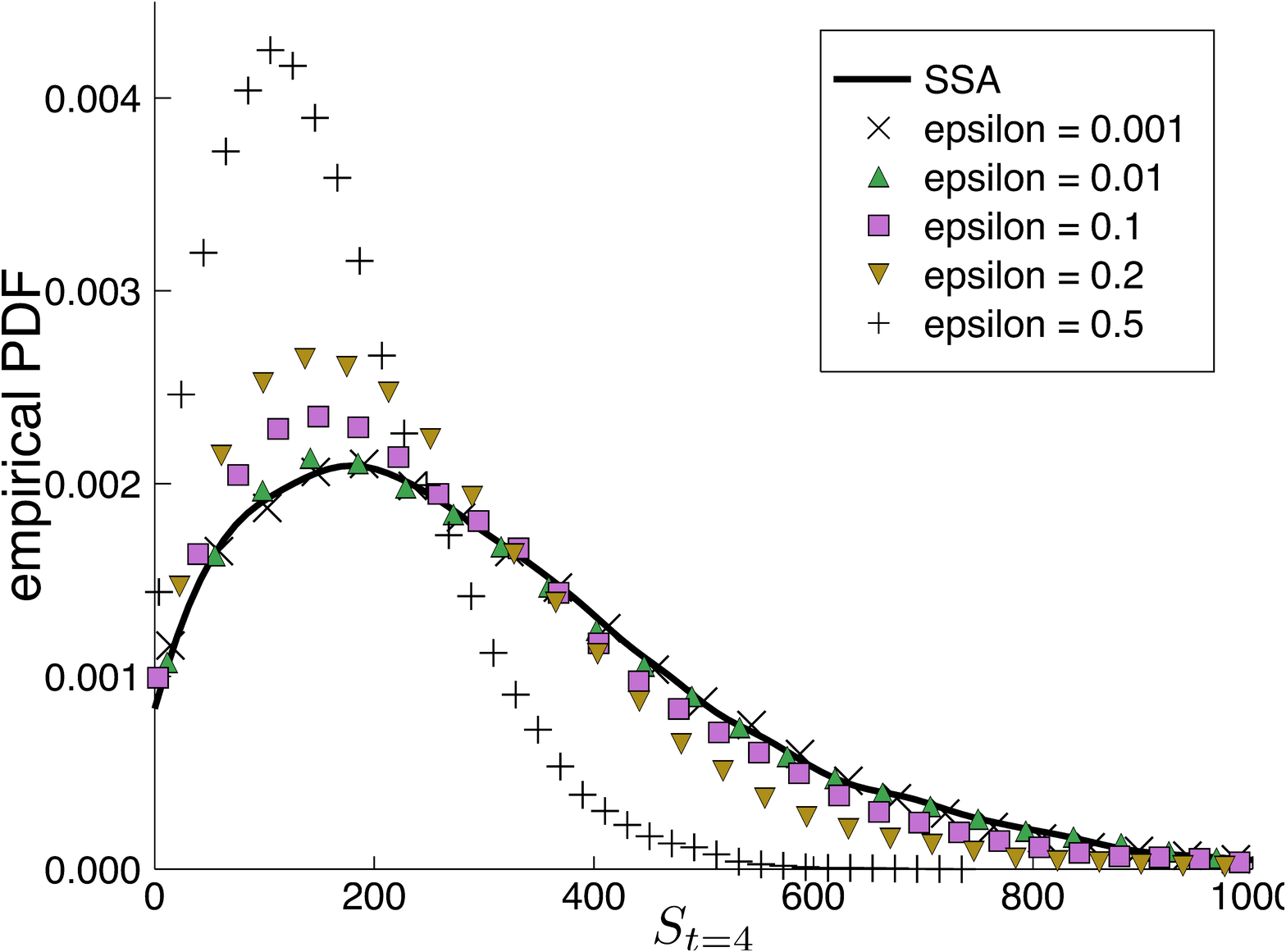}
    \caption{}
  \end{subfigure}
  %%%%% Panel B %%%%%
  \begin{subfigure}[b]{0.49\textwidth}
    \includegraphics[width=\textwidth]{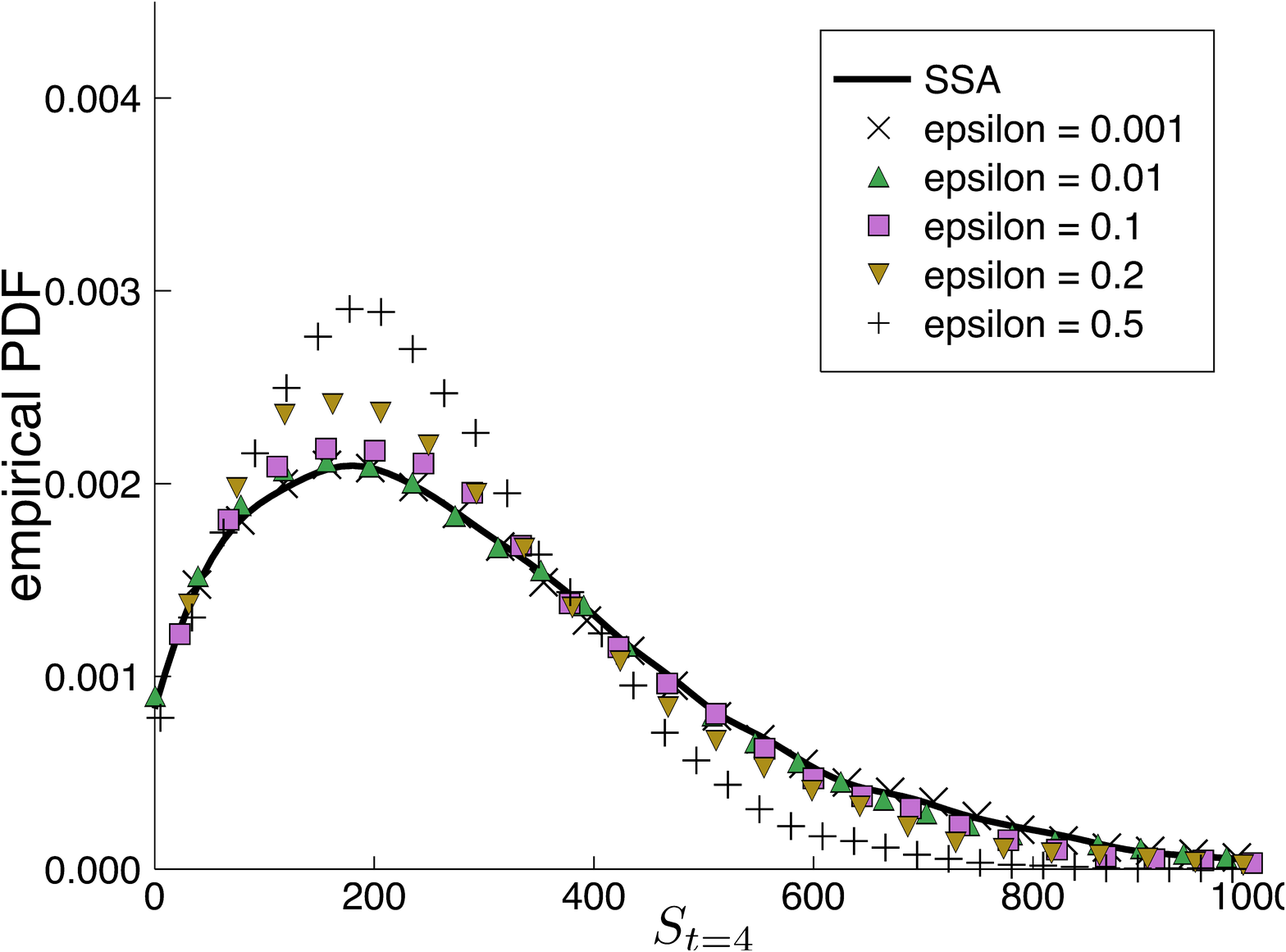}
    \caption{}
  \end{subfigure}
  \caption{The distribution of $S$ in Kendall's process at $t = 4$ computed from $10^5$ realizations using (a) OTL and (b) SAL.
  SSA is used to derive an exact estimate of the distribution for comparison and is shown in black.
  The colored lines represent the resulting distribution under different values of $\epsilon$.
  Conservative values are generally safe, but may increase the clock time performance of a specific algorithm on a non-trivial model.
  In this example, the time to generate all $10^{5}$ realizations for SSA is approximately $5$ seconds.
  The running times for OTL are (in order of increasing $\epsilon$ value): $6$, $3$, $0.4$, $0.2$, and $0.1$ seconds.
  Similarly, the running times for SAL are: $9$, $6$, $0.2$, $0.1$, and $0.05$ seconds.\label{fig:kendall_tradeoff}}
\end{figure}

In addition to tracking the mean and variance, \pkg{BioSimulator.jl} enables the display of the full distribution of species counts.
In this way, one can quantify the frequency of rare extinction events.
Figure~\ref{fig:kendall_plots} summarizes results for the mean trajectory and distribution of species $S$ in Kendall's process.    
At $t=4$, the average population is $300$, and in approximately $0.9$ percent of the simulations, the species has gone extinct by this time point.
Figure~\ref{fig:kendall_tradeoff} illustrates the trade-off in selecting large $\epsilon$ values in $\tau$-leaping methods, namely OTL and SAL.
As $\epsilon$ decreases towards $0$, the distribution of $S$ at $t = 4$ approaches the exact statistical results from the SSA.
Note that the smaller $\epsilon$ values may force these $\tau$-leaping algorithms to perform slower than SSA because the proposed leap sizes become smaller than an SSA update.
This means that each update wastes time computing a $\tau$ leap that is often rejected.
On the other hand, large $\epsilon$ values tend to increase the number of bad leaps.
In this case, $\tau$-leaping wastes time contracting the leap size until a suitable update emerges.

\begin{example}Michaelis-Menten enzyme kinetics\end{example}

\begin{figure}[hp]
	\centering
    %%%%% Panel A %%%%%
    \begin{subfigure}[b]{0.49\textwidth}
        \includegraphics[width=\textwidth]{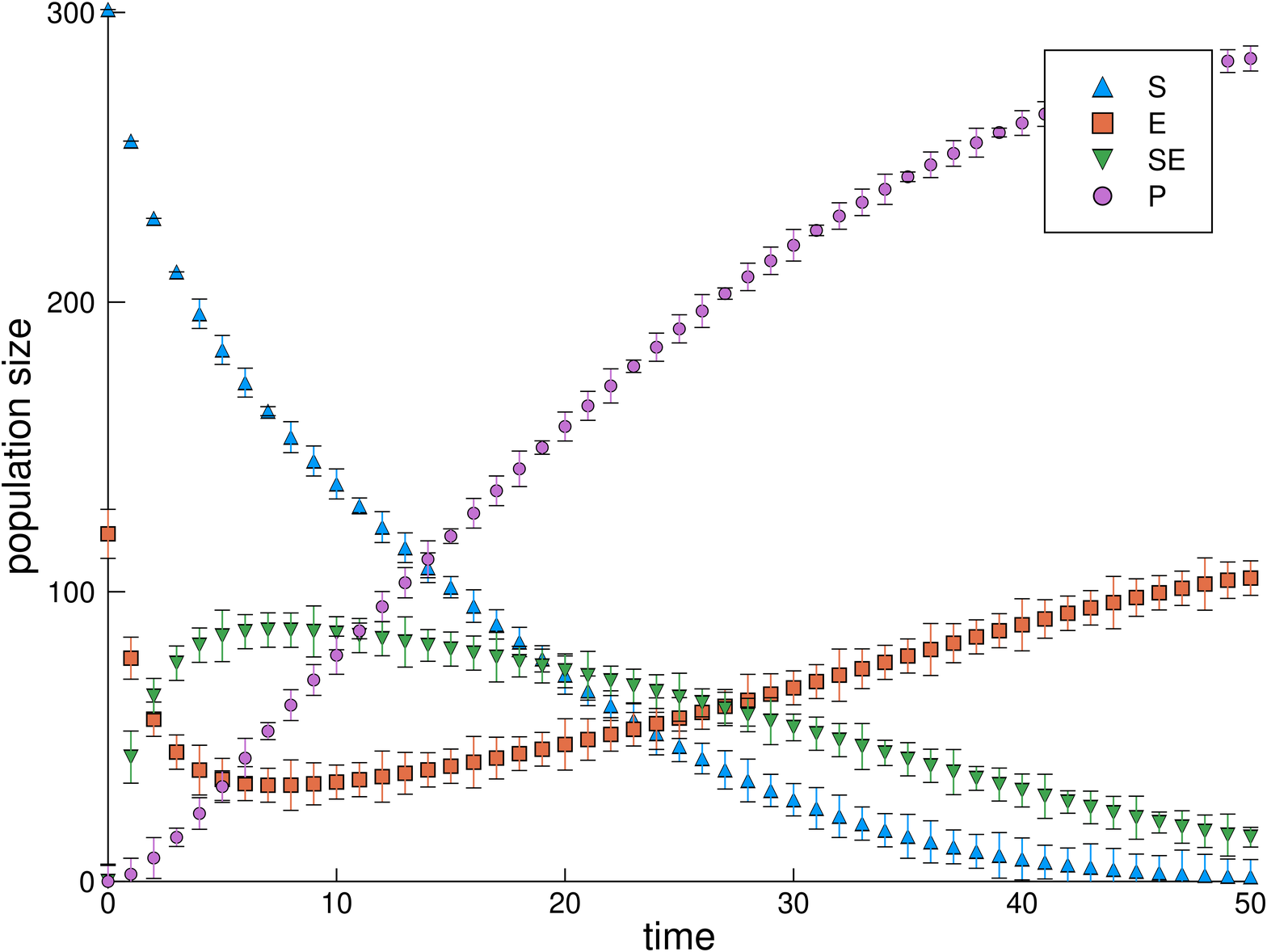}
        \caption{}
    \end{subfigure}
    %%%%% Panel B %%%%%
    \begin{subfigure}[b]{0.49\textwidth}
        \includegraphics[width=\textwidth]{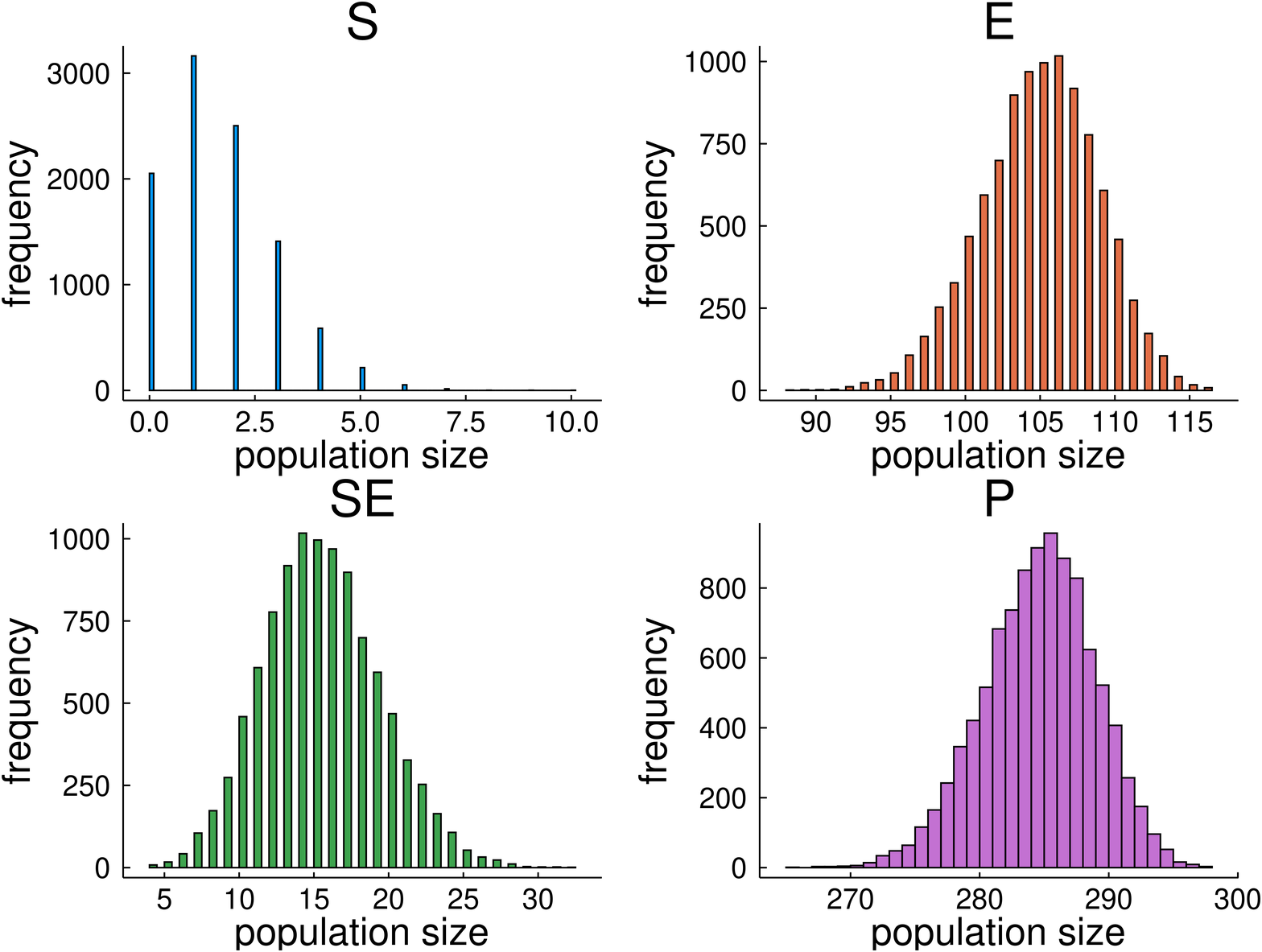}
        \caption{}
    \end{subfigure}
	\caption{(a) Mean trajectory of each species in the Michaelis-Menten model using $10^4$ realizations. The region representing one standard deviation from the mean is small and suggests this network is not dominated by noise. (b) Population distributions for each species at $t=50$ generated from $10^4$ realizations.\label{fig:mmek}}
\end{figure}

Our next example, Michaelis-Menten enzyme kinetics, involves a combination of first- and second-order reactions.
The system consists of a substrate S, an enzyme E, the substrate-enzyme complex SE, and a product P.
The three reactions connecting them
\begin{equation*}
\left\{
    \begin{aligned}
        S + E & \rightarrow SE &\text{Binding}\\
        SE    & \rightarrow S + E &\text{Dissociation}\\
        SE    & \rightarrow P + E
        &\text{Conversion}
    \end{aligned}
\right.
\end{equation*}
represent binding of the substrate to an enzyme, dissociation of the substrate-enzyme complex, and conversion of the substrate into a product.
These reactions have rates \( a_{1} \), \( a_{2} \), and \( a_{3} \), respectively.
The following code simulates this system under the initial conditions $S = 301$, $E = 130$, $SE = P = 0$ and rate constants $a_{1} = 0.00166$, $a_{2} = 0.0001$, and $a_{3} = 0.1$.
\begin{minted}[fontsize=\scriptsize]{julia}
    using BioSimulator

    model = Network("Michaelis-Menten")

    model <= Species("S", 301)
    model <= Species("E", 130)
    model <= Species("SE",  0)
    model <= Species("P",   0)
    
    model <= Reaction("dimerization", 0.00166, "S + E --> SE")
    model <= Reaction("dissociation", 0.0001,  "SE --> S + E")
    model <= Reaction("conversion",   0.1,     "SE --> P + E")

    result = simulate(model, StepAnticipation(), time = 50.0, epochs = 10_000, trials = 1_000)
\end{minted}
Figure~\ref{fig:petrinets}~(b) provides a Petri net representation of the model.
Figure~\ref{fig:mmek}~(a) demonstrates mean trajectories and standard deviations for each of the reactant species over time.
These mean trajectories match the dynamics predicted by a deterministic model.
Figure~\ref{fig:mmek}~(b) shows the full distribution of species counts after $t=50$ time steps. The substrate is typically exhausted at $t=50$.

\begin{example}Auto-regulatory genetic network\end{example}

The influence of noise at the cellular level is difficult to capture in deterministic models.
Stochastic simulation is appropriate for the study of regulatory mechanisms in genetics, where key species may be present in low numbers.
Figure~\ref{fig:petrinets}~(c) is an example of a simplified \textit{negative} auto-regulation network for a single gene, in the sense that the protein represses its own transcription.

There are eight possible reactions: (1) gene transcription into RNA, (2) translation of the protein, (3) dimerization of the protein with itself, (4) dissociation of the protein dimer, (5) binding to the gene, (6) unbinding from the gene, (7) RNA degradation, and (8) protein degradation.
There are five species to track --- the free copies of the gene, transcribed RNA, protein molecules, dimer molecules, and blocked copies of the gene.
The model is easily implemented in \pkg{BioSimulator.jl}:
\begin{minted}[fontsize=\scriptsize,samepage]{julia}
    using BioSimulator

    model <= Network("negative auto-regulation")

    model <= Species("gene", 10) # assume 10 copies of the gene are present
    model <= Species("RNA", 0)   # transcribed from the underlying gene
    model <= Species("P", 0)     # protein
    model <= Species("P2", 0)    # protein dimer
    model <= Species("P2_gene")  # gene repression

    model <= Reaction("transcription",              0.01, "gene --> gene + RNA")
    model <= Reaction("translation",                10.0, "RNA --> RNA + P")
    model <= Reaction("dimerization",               1.0,  "P + P --> P2")
    model <= Reaction("dissociation",               1.0,  "P2 --> P + P")
    model <= Reaction("repression binding",         1.0,  "gene + P2 --> P2_gene")
    model <= Reaction("reverse repression binding", 10.0, "P2_gene --> gene + P2")
    model <= Reaction("RNA degradation",            0.1,  "RNA --> 0")
    model <= Reaction("protein degradation",        0.01, "P --> 0")

    result = simulate(model, StepAnticipation(), Val(:full), time = 500.0, trials = 100)
\end{minted}
RNA typically has a limited lifetime. Thus, the per particle reaction rates governing protein production are balanced to favor translation events following transcription.
Moreover, the reaction rates for dimerization and dissociation reflect an assumption that the protein favors neither the monomer nor the dimer configuration.
\begin{figure}[ht]
    \centering
    %%%%% Panel A %%%%%
    \begin{subfigure}[b]{0.49\textwidth}
        \includegraphics[width=\textwidth]{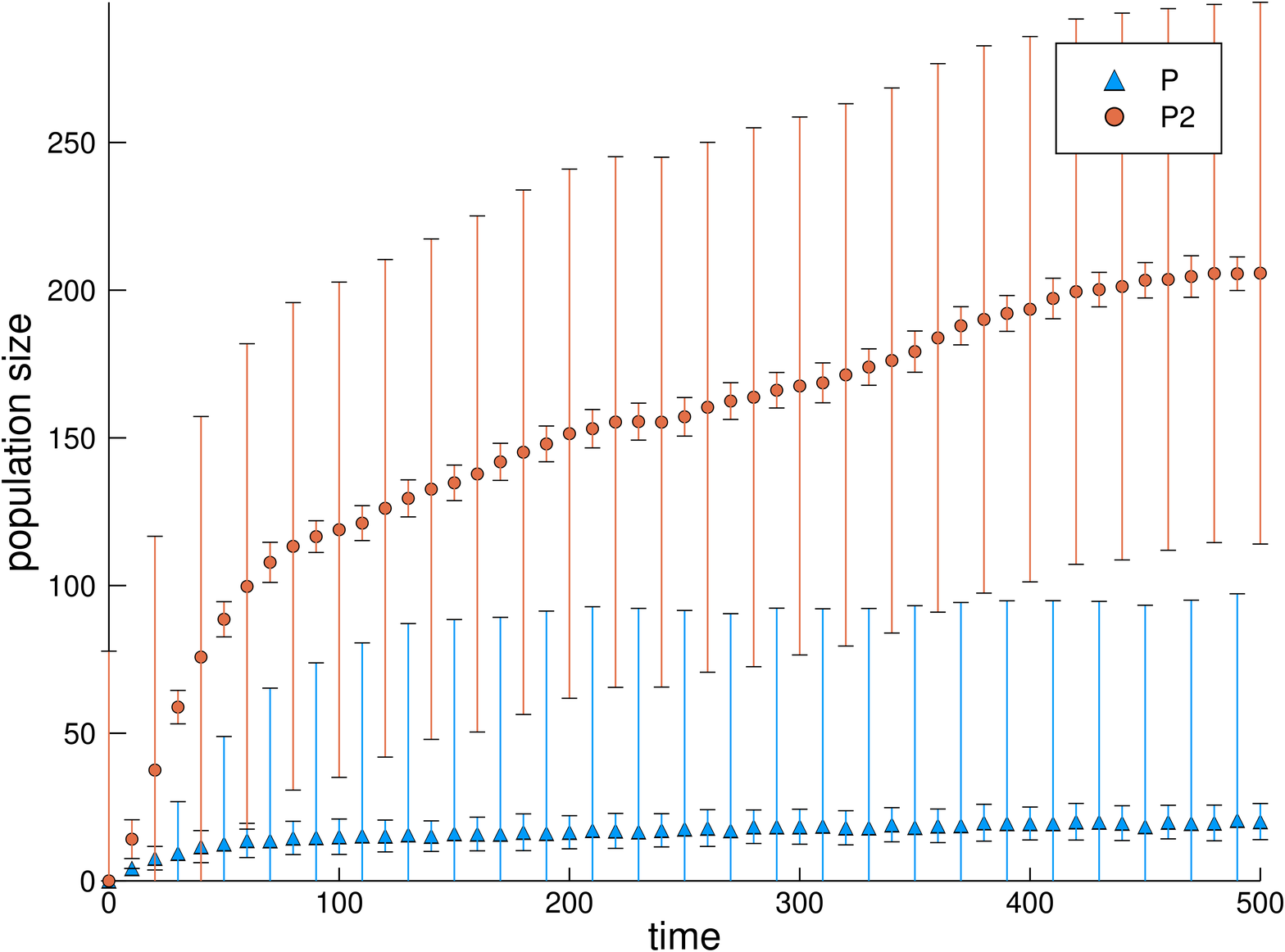}
        \caption{}
    \end{subfigure}
    %%%%% Panel B %%%%%
    \begin{subfigure}[b]{0.49\textwidth}
        \includegraphics[width=\textwidth]{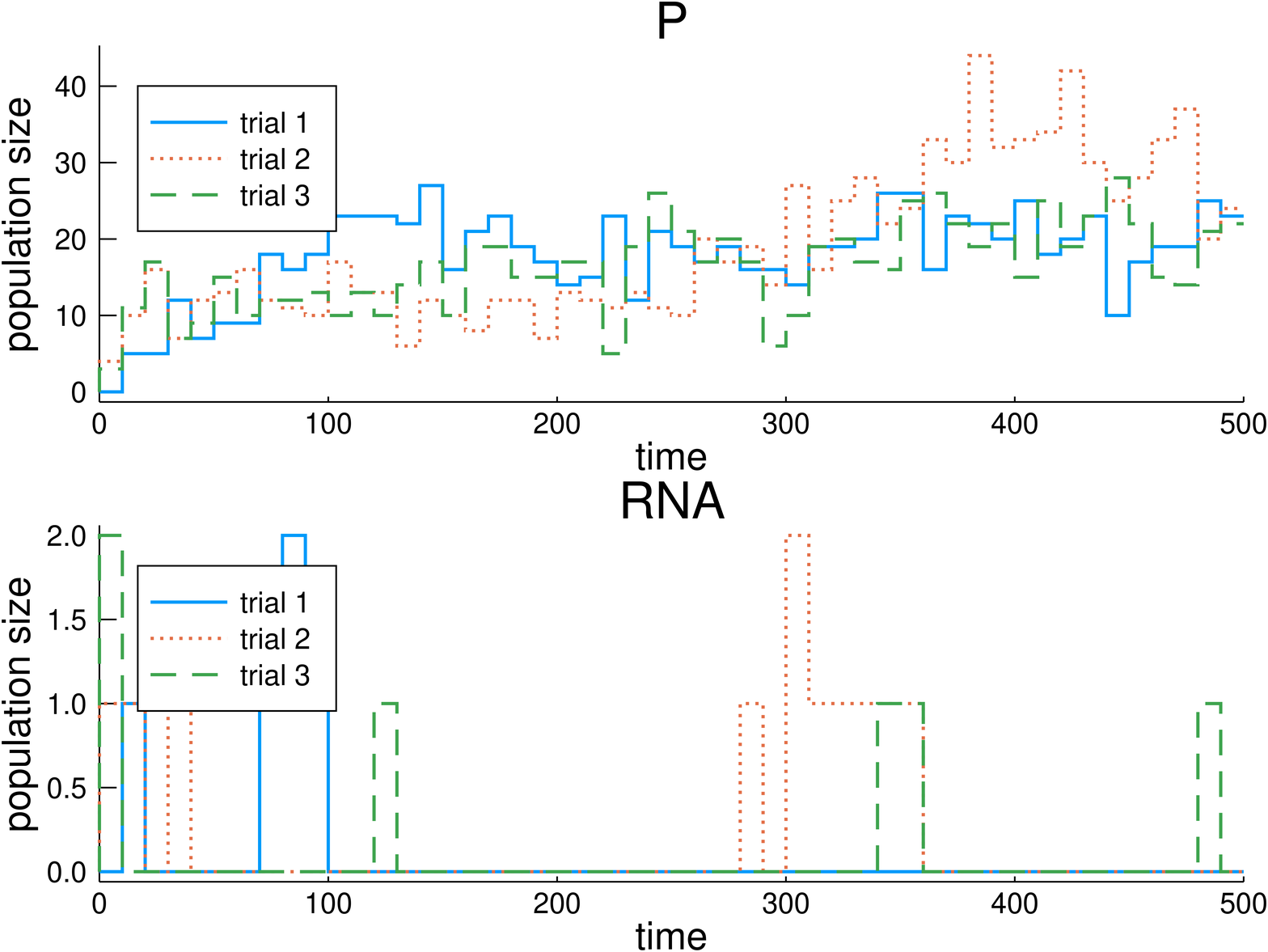}
        \caption{}
    \end{subfigure}
    \caption{(a) Mean trajectories for the protein and dimer in a simple auto-regulatory gene network. The bars represent $1$ standard deviation away from the mean.
    (b) Full sample paths for the protein (top) and RNA (bottom). Note that the RNA sample path is sensitive to the number of epochs if one were using the fixed-interval option. Using a large window may fail to capture peaks \textit{in the output}. This subtlety is important for qualitative model assessment. \label{fig:autoreg}}
\end{figure}
Figure~\ref{fig:autoreg}~(a) compares the mean behavior of the protein and the dimer over time with results from a deterministic model.
Plotting individual trajectories for the protein level reveals strong stochastic fluctuations driven by the relative rarity of RNA\@.
Figure~\ref{fig:autoreg}~(b) highlights the influence of noise in the system's dynamics.

\subsection{Algorithm comparison}

Here we compare the performance of \pkg{BioSimulator.jl}'s algorithms across the three examples from the previous sections.
Each model is simulated using the same model parameters outlined in the previous examples.
Each simulation task involves generating $100$ realizations.
In the case of fixed-interval output, $1000$ epochs are used.
The $\tau$-leaping algorithms use the default values $\epsilon = 0.03$, $\delta = 2$, and $\beta = 0.75$.

Table~\ref{tab:biosimulator-benchmarks} records the \textit{averaged} clock times reported by the \code{@benchmark} macro from \proglang{Julia}'s \pkg{BenchmarkTools.jl} package.
Each simulation task is allotted $2$ minutes to generate $100$ samples of the running time.
To be explicit, this means that the following command is executed \textit{at most} $100$ times:
\begin{minted}[fontsize=\scriptsize,samepage]{julia}
  simulate(model, algname, output_option, time = tfinal, epochs = 1000, trials = 100)
\end{minted}
Thus, these benchmarks reflect both the cost of generating a single realization, the efficiency of completing a typical simulation task, and the economy of choosing fixed-interval versus full simulation output.
Results are based on a MacBook Pro with 2 GHz Intel Core i7 (4 cores) and 16 GB of RAM running macOS High Sierra 10.13.3.
\begin{table}[!htbp]
  \centering
  \ra{1.3}
  \begin{tabular}{@{}cccc@{}}
    \toprule
    Algorithm & Kendall's process & Michaelis-Menten & Auto-regulation \\
    \midrule
    \multirow{2}{*}{SSA}
      & $1.04\ (1.03 - 1.05)$ & $0.783\ (0.788 - 0.808)$ & $426\  (424 - 432)$ \\
      & $1.38\ (1.37 - 1.40)$ & $0.918\ (0.909 - 0.936)$ & $1809\ (1734 - 1860)$ \\
    \midrule
    \multirow{2}{*}{FRM}
      & $1.04\ (1.03 - 1.05)$ & $0.863\ (0.853 - 0.874)$ & $464\  (462 - 468)$ \\
      & $1.35\ (1.33 - 1.37)$ & $0.983\ (0.973 - 1.000)$ & $1597\ (1408 - 1645)$ \\
    \midrule
    \multirow{2}{*}{NRM}
      & $2.72\ (2.71 - 2.75)$ & $3.11\ (3.08 - 3.15)$ & $1589\ (1581 - 1597)$   \\
      & $2.99\ (2.96 - 3.02)$ & $3.28\ (3.25 - 3.33)$ & $2724\ (2553 - 2775)$ \\
    \midrule
    \multirow{2}{*}{ODM}
      & $1.02\ (1.01 - 1.03)$ & $0.839\ (0.820 - 0.853)$ & $409\  (407 - 412)$ \\
      & $1.34\ (1.32 - 1.37)$ & $0.952\ (0.946 - 0.967)$ & $1648\ (1437 - 1721)$ \\
    \midrule
    \multirow{2}{*}{OTL}
      & $0.722\ (0.715 - 0.726)$ & $1.50\ (1.49 - 1.52)$ & $1867\ (1863 - 1877)$  \\
      & $0.754\ (0.751 - 0.760)$ & $1.64\ (1.63 - 1.67)$ & $3038\ (2845 - 3087)$ \\
    \midrule
    \multirow{2}{*}{SAL}
      & $0.596\ (0.588 - 0.599)$ & $0.989\ (0.985 - 0.996)$ & $1422\ (1415 - 1441)$  \\
      & $0.607\ (0.604 - 0.612)$ & $0.957\ (0.954 - 0.967)$ & $2578\ (2398 - 2631)$ \\
    \bottomrule
  \end{tabular}
  \caption{Median runtimes and interquantile ranges (in milliseconds) for Example 1, Example 2, and Example 3 based on $1000$ samples of the \code{simulate} command.
  Each simulation task generates $10$ realizations of the underlying stochastic model.
  For example, the median time to generate $10$ realizations of \textit{Kendall's process} using SSA is $1.04$ milliseconds.
  The first row in a cell is the benchmark result using the \code{Val(:fixed)} option for fixed-interval output.
  The second row indicates the timing result using the \code{Val(:full)} option for full simulation output.
  \label{tab:biosimulator-benchmarks}}
\end{table}

The NRM performs the worst across the selected model.
However, we note that \pkg{BioSimulator.jl} uses the a priority queue implementation from the \pkg{DataStructures.jl} package which may not be optimal for the method's specific demands.
Moreover, we take a na\"ive approach to building dependency graphs.
The overhead from \pkg{BioSimulator.jl}'s dependency graphs is sufficiently large that in some cases most of the simulation is spent on this step.
This warrants further review of our NRM implementation, although we expect an improved implementation to perform similarly to the FRM.
Only the auto-regulation model may show improved performance for the NRM over SSA because the reaction channels for that model are not tightly coupled.
Similarly, our ODM implementation may be inefficient based on its performance on the birth-death-immigration process.

Both $\tau$-leaping algorithms perform poorly on the auto-regulation model.
This is to be expected because neither OTL nor SAL handle separate time scales.
Figure~\ref{fig:autoreg}b shows that RNA transcription becomes a rare event as the protein dimer population grows.
Similarly, the free and blocked gene copies are always present in relatively small quantities compared to the protein and protein dimer populations.
Thus, protein dimerization and dissociation become the dominant reaction channels further along the time axis.
These two reactions drive the dynamics of the system and therefore heavily influence the $\tau$-leap selection procedure.
A natural consequence is that $\tau$-leaping methods have an increased likelihood of violating the leap condition in this regime and therefore bias the simulation toward infeasible states.
This assertion is easily verified using the \code{track_stats = true} option and checking the number of negative excursions reported by \pkg{BioSimulator.jl}.
In fact, the leap condition violations are egregious to the point that the recovery mechanism requires thinning the leap several times per negative excursion.
This suggests that reducing the leap size via the $\epsilon$ control parameter is likely to reduce the algorithm to SSA because leaps eventually become smaller than the expected Gillespie updates.

\subsection{Software comparison}

We compare \pkg{BioSimulator.jl}'s features and performance against three software packages: \pkg{StochPy}, \pkg{StochKit2}, and \pkg{Gillespie.jl}.
The tools in the space of stochastic simulation are manifold and vary in their features and applications.
Naturally, there are many omissions.
Each software package is selected for its similarity to \pkg{BioSimulator.jl}.
Specifically, these tools are domain-independent, general purpose Gillespie-like simulators.
Table~\ref{table:soft-compare} summarizes the differences and similarities between these software tools.
\begin{table}
  \centering
  \ra{1.3}
  \begin{tabular}{@{}rccccccc@{}}
    \toprule
    & \pkg{StochPy} & \pkg{StochKit2} & \pkg{Gillespie.jl}     & \pkg{BioSimulator.jl} \\
    \midrule
    \textbf{Software characteristics}: & & & & \\
    Language & Python  & C++       & Julia            & Julia \\
    Open-source & Yes & Yes & Yes & Yes \\
    GUI & No & via StochSS & No & No \\
    Jupyter integration & Yes & No & Yes & Yes \\
    Performance & Fast & Faster & Faster & Fastest \\
    \textbf{Interactivity}: &     &             &     & \\
    Model editor            & No  & via StochSS & No  & Yes \\
    Simulation interface    & Yes & via StochSS & Yes & Yes \\
    Plotting         & Yes & Limited & No & Yes \\
    \textbf{Simulation features}: & & & & \\
    Fixed-interval output & via StochKit2 & Yes & No & Yes \\
    Full output & Yes & Yes & Yes & Yes \\
    SBML support & Yes & Yes & No & No \\
    Readable input & Yes & No & No & Yes \\
    Parallelism & No & Yes & No & Yes \\
    \bottomrule
  \end{tabular}
  \caption{A summary of features across \pkg{StochPy}, \pkg{StochKit2}, \pkg{Gillespie.jl}, and \pkg{BioSimulator.jl}. Jupyter notebooks are human-readable documents that combine code, text, and figures into a single interactive report.}
  \label{table:soft-compare}
\end{table}

In addition, we compare each software package on simulation speed across five different models:
\begin{enumerate}[label=(\alph*)]
  \item \textit{Kendall's process} permits an analytic mean trajectory.
  This model serves as an easy speed test and a correctness check.

  \item The \textit{Michaelis-Menten enzyme kinetics} model provides a second simple speed test with multiple species.

  \item The \textit{auto-regulation} model serves a benchmark against an interesting biological application of stochastic simulation.

  %%%% cite a paper for the model %%%%
  \item The \textit{dimer-decay} model is an example of a small stiff system.
  It is featured in the literature as a $\tau$-leaping benchmark \citep{efficient,slow-scale-tau}.

  %%%% cite a paper for the model %%%%
  \item A \textit{yeast} model from biology is another interesting stiff system used here as a benchmark. \citep{chou2008,drawert2010,yeast-biology}.
\end{enumerate}
Initial conditions and other model parameters are deferred to supplementary information (Section A.1, Tables S1 - S5).
Table~\ref{tab:benchmarks} reports the benchmark results and includes the \textit{simulation} parameters.

\begin{landscape}
\begin{table}[!htbp]
\centering
\centering
\ra{1.3}
\begin{tabular}{@{}crrrrr@{}}
    \toprule
    Software & \multicolumn{1}{c}{\pkg{StochPy}} & \multicolumn{1}{c}{\pkg{StochKit2}} & \multicolumn{1}{c}{\pkg{Gillespie.jl}} & \multicolumn{2}{c}{\pkg{BioSimulator.jl}} \\
    Method   & \multicolumn{1}{c}{SSA}           & \multicolumn{1}{c}{SSA (dep. graph)}             & \multicolumn{1}{c}{SSA}                & \multicolumn{2}{c}{SSA} \\
    Mode     & \multicolumn{1}{c}{Serial}        & \multicolumn{1}{c}{Parallel}        & \multicolumn{1}{c}{Serial}             & \multicolumn{1}{c}{Serial} & \multicolumn{1}{c}{Parallel} \\
    \midrule % biosimulator is currently in ms
    \multirow{2}{*}{Kendall's process}
      &                    & $174.31\ (173.81 - 174.95)$ &                       & $1.04\ (1.03 - 1.05)$ & $0.56\ (0.48 - 0.66)$ \\
      & $257\ (223 - 296)$ &                             & $1285\ (1280 - 1293)$ & $1.38\ (1.37 - 1.40)$ & $0.79\ (0.68 - 0.93)$ \\
    \multirow{2}{*}{Michaelis-Menten}
      &                    & $210.42\ (209.89 - 211.14)$ &                       & $0.783\ (0.788 - 0.808)$ & $0.45\ (0.44 - 0.46)$ \\
      & $214\ (213 - 216)$ &                             & $1.15\ (1.12 - 1.21)$ & $0.918\ (0.909 - 0.936)$ & $0.47\ (0.46 - 0.48)$ \\
    \multirow{2}{*}{Auto-regulation}
      &                                      & $432\ (425 - 445)$ &                    & $426\  (424 - 432)$ & $135\ (124 - 149)$ \\
      & $8.23\ (7.87 - 8.75); \times 10^{4}$ &                    & $824\ (762 - 834)$ & $1809\ (1734 - 1860)$ & $211\ (193 - 240)$ \\
    \multirow{2}{*}{Dimer-decay}
      &                    & $216.98\ (216.52 - 217.68)$ &                       & $1.76\ (1.75 - 1.79)$ & $0.81\ (0.80 - 0.82)$ \\
      & $483\ (480 - 488)$ &                             & $2.68\ (2.66 - 2.84)$ & $2.39\ (2.37 - 2.42)$ & $1.03\ (1.02 - 1.05)$ \\
    \multirow{2}{*}{Yeast}
      &                    & $260\ (254 - 267)$ &                       & $2.56\ (2.53 - 2.63)$ & $1.38\ (1.10 - 1.42)$ \\
      & $469\ (465 - 474)$ &                    & $2.25\ (2.21 - 2.63)$ & $2.92\ (2.89 - 2.99)$ & $1.24\ (1.20 - 1.27)$ \\
    \bottomrule
\end{tabular}
\caption{Median runtimes and interquantile ranges for \pkg{StochPy}, \pkg{StochKit2}, \pkg{Gillespie.jl}, and \pkg{BioSimulator} across selected models based on $1000$ samples, reported in milliseconds (ms).
Each sample measured the time to generate $10$ realizations of a given stochastic process.
Results using fixed-interval and fixed output options are recorded in the first and second rows of each cell, respectively.
Each simulation tool is used with its default settings.
For example, \pkg{StochKit2} automatically parallelizes simulation tasks involving multiple realizations and uses a dependency graph by default.
We note that both \pkg{StochKit2} and \pkg{BioSimulator.jl} used $8$ threads for the parallel simulation benchmarks.
Direct comparisons based on these results are not possible, in the sense that slower performance does not necessarily indicate a particular tool is poorly implemented.
Rather, our results reflect natural trade-offs in optimizing software for particular goals.
\textbf{Note}: The \pkg{StochPy} benchmark on the \textit{auto-regulation} model is based on only $100$ samples.
\label{tab:benchmarks}}
\end{table}
\end{landscape}

\subsection{Package availability}

\pkg{BioSimulator.jl} is available on GitHub (\url{https://github.com/alanderos91/BioSimulator.jl}).
All source code is readily available to view, download, and distribute under the MIT license.
We also maintain a documentation manual via GitHub Pages (\url{https://alanderos91.github.io/BioSimulator.jl/stable/}).

\section{Discussion}

\pkg{BioSimulator.jl} simplifies interactive stochastic modeling by virtue of being contained within a single programming language.
Every aspect of the modeling workflow --- model specification, simulation, and analysis --- is handled by \proglang{Julia} and its type system.
Here we discuss these three aspects of our simulation software and compare its features and timing benchmarks with other packages.

\pkg{StochPy} and \pkg{BioSimulator.jl} are mainly interactive simulation tools meant to be used in Read-Eval-Print Loop (REPL) environments and Jupyter notebooks, but they stand out in that they can also be used as libraries like \pkg{StochKit2}.
\pkg{BioSimulator.jl}'s model specification interface is flexible enough to allow a user to build up a model using for loops and other language features.
\pkg{StochPy} and \pkg{StochKit2} mainly rely on input files whose main advantage is model sharing.
Model editing interfaces can be built around standardized formats and indeed such tools exist; \pkg{StochSS} includes a graphical user interface that feeds into \pkg{StochKit2}.
The strength of \pkg{BioSimulator.jl}'s model interface is the flexibility afforded by \proglang{Julia}'s features as it facilitates rapid model prototyping.

An additional benefit of our implementation is that models can be packaged into functions by defining a wrapper.
This allows one to share models through \code{.jl} files that provide the model through a function call that specifies parameters and initial conditions.
A model author has full control over what parameters ought to be exposed to a user by designing the function signature appropriately.
As an example, recall the model definition for Kendall's process:
\begin{minted}[fontsize=\scriptsize,samepage]{julia}
  function birth_death_process(S; birth_rate = 2.0, death_rate = 1.0, immigration_rate = 0.5)
    model = Network("Kendall's Process")

    model <= Species("S", S)

    model <= Reaction("Birth",       birth_rate,       "S --> S + S")
    model <= Reaction("Death",       death_rate,       "S --> 0")
    model <= Reaction("Immigration", immigration_rate, "0 --> S")

    return model
  end
\end{minted}
Here \code{birth_death_process} is a function that requires an argument \code{S} to specify an initial condition.
The variables \code{birth_rate}, \code{death_rate}, and \code{immigration_rate} are optional and have default values.
Calling \code{birth_death_process(5)} builds up the model with $S_{0} = 5$ and the default reaction rate constants.

Many models are implemented using data standards independent of software and programming languages.
One notable standard is the Systems Biology Markup Language (SBML) \citep{libsbml}.
SBML addresses crucial details such as parameter units, compartment sizes, and kinetic rate laws in a standardized format.
An interface to SBML is required in order to connect \pkg{BioSimulator.jl} to existing software, enable thorough comparisons, and facilitate model sharing in a standardized way.
\pkg{StochPy} and \pkg{StochKit2} stand out in this regard because these tools provide an interface to SBML.
We anticipate adopting the SBML standard in future versions of our software.

\pkg{StochKit2} is primarily a command-line tool and a software development package.
This makes \pkg{StochKit2} highly reusable since other programs, such as model editors or simulation packages, can interface with it.
In fact, one of \pkg{StochPy}'s features is its ability to call \pkg{StochKit2} for tau-leaping.
The main commands are \code{ssa} and \code{tau_leaping}.
At the start of a simulation job, \pkg{StochKit2} performs an analysis to select a suitable algorithm based on the model structure.
For example, using the \code{ssa} command will run some variation of the exact Gillespie algorithm based on model structure.
The main drawback is the command-line interface, which may be off-putting to inexperienced users.

Overall, \pkg{Gillespie.jl} is a fast stochastic simulation tool for visualizing results and computing quantities of interest, but it requires a modest programming effort.
Without a model editor, it requires the user to specify the net change increments $\mathbf{v}^{j}$ and propensity functions $r_{j}(\mathbf{x})$ for each reaction $j$.
This approach places a burden on users, especially those unfamiliar with \proglang{Julia} or any similar programming language.
A benefit of this interface is that non-mass action propensities are automatically supported since a user must hard-code these for the software.
Unlike the other tools, \pkg{Gillespie.jl} does not yet have an interface for running multiple simulations, and so collecting the results from independent simulations requires some effort from a user.
While the main simulation functions are similar, \pkg{Gillespie.jl} and \pkg{BioSimulator.jl} have the advantage of \proglang{Julia}'s type system and multiple-dispatch.
When used correctly, these two language features allow one to write powerful abstractions around scientific computing problems that \proglang{Julia} leverages to generate highly optimized instructions.
The \pkg{JuMP.jl} and \pkg{DifferentialEquations.jl} packages are exemplars in this regard and serve as a testament to \proglang{Julia}'s advantages \citep{juMP2017, diffeq2017}.

The benchmarking results from Table~\ref{tab:benchmarks} show that \pkg{BioSimulator.jl} is competitive with \pkg{StochKit2}.
While we are pleased with \pkg{BioSimulator.jl}'s performance, one can only speculate on the precise meaning of these benchmarks without an intimate understanding of implementation details.
\proglang{Julia} provides a theoretical competitive performance edge for \pkg{Gillespie.jl} and \pkg{BioSimulator.jl} over the \proglang{Python} software.
\pkg{StochPy}'s timings may be slower due to the fact that it records a greater amount of information.
In addition to recording the state vector after each step, \pkg{StochPy} offers the option to store reaction channel information such as propensity values and event waiting times.
While fixed-interval output provides \pkg{StochKit2} and \pkg{BioSimulator.jl} with a slight performance boost, the results between \pkg{StochPy} and \pkg{BioSimulator.jl} using the full output option suggest that the performance gap is in fact wide.

\pkg{StochKit2} and \pkg{BioSimulator.jl} each support running parallel simulations.
\pkg{StochKit2} automates this feature; the software defaults to parallelism if a user's machine supports it.
However, \pkg{BioSimulator.jl}'s implementation is nearly automatic thanks to \proglang{Julia}'s abstractions for parallelism.
Table~\ref{tab:benchmarks} shows that \pkg{BioSimulator.jl} is an order of magnitude faster on each example in the serial case, except for the auto-regulation model.
This warrants further investigation.

All four packages support varying degrees of simulation output analysis.
\pkg{Gillespie.jl} and \pkg{BioSimulator.jl} offer automatic time series visualization.
We improve upon \pkg{Gillespie.jl} by providing helper functions for additional visualization tools, such as histograms and mean trajectories, as well as integration with the \pkg{Plots.jl} ecosystem.
Our goal is to emulate \pkg{StochKit2} / \pkg{StochSS} and \pkg{StochPy} in their support for more sophisticated analysis tools.

\section{Conclusion}
In a biological system, interacting feedback loops can make mathematical analysis intractable and create challenges in choosing an optimal set of experiments to probe system behavior.
Combining experiments and stochastic simulation of complex biological systems promotes model validation and the design of promising experiments.
The user-friendly nature of \pkg{BioSimulator.jl} encourages the use of stochastic simulation, eliminates effort spent on modifying simulation code, reduces errors during model specification, and allows visualization of system interactions via Petri Nets. 
Tracking trajectories and distributions of interacting species over time helps modelers decide between deterministic and stochastic models.
Future developments of \pkg{BioSimulator.jl} include
\begin{itemize}
  \item support for non-mass action kinetics,

  \item adopting SBML as an input format,

  \item extending the SAL algorithm implementation using higher order Taylor expansions,

  \item implementing additional exact and approximate simulation algorithms from the literature,

  \item incorporating spatial effects, and

  \item implementing hybrid methods that integrate stochastic simulation with deterministic modeling.
\end{itemize}

The \proglang{Julia} language provides an ideal environment for this purpose.
Its syntax allows one to write mathematical code in a natural way and facilitates fast prototyping.
In particular, the integration with the \pkg{IJulia} package encourages code sharing and reproducible work.
Furthermore, its interactive environment is well-suited for novice programmers.
The package ecosystem in \proglang{Julia} provides software for visualization, statistical analysis, optimization, differential equations, and probability distributions.
Emerging computational tools in \proglang{Julia} can only increase \pkg{BioSimulator.jl}'s strengths over time.

%%%%% Acknowledgements %%%%%
\section*{Acknowledgments}

We thank Kevin Tieu for testing an early version of our software.
This work was funded by NCATS Grant KL2TR000122.
A.L.\ and T.S.\ were funded by the NIH Training Grant in Genomic Analysis and Interpretation T32HG002536.
K.L.K. was supported by grant 5R01HL135156-02S1, the UCSF Bakar Computational Health Sciences Institute, and the UC Berkeley Institute for Data Sciences as part of the Moore-Sloan Data Sciences Environment initiative.

%%%%% bibliography %%%%%
\bibliography{references}
\bibliographystyle{plainnat}
\end{document}

\section*{Supplementary Information}

\subsection{Benchmark models}

\begin{model}Kendall's process\end{model}
  \begin{equation*}
    \left\{
    \begin{aligned}
      S & \rightarrow S + S &\text{Birth}\\
      S & \rightarrow 0     &\text{Death}\\
      0 & \rightarrow S     &\text{Immigration}
    \end{aligned}
    \right.
  \end{equation*}

  \begin{table}[!htbp]
    \centering
    \ra{1.3}
    \begin{tabular}{@{}crr@{}}
      \toprule
      Reaction $j$ & Rate constant $a_{j}$ & Propensity $r_{j}(\mathbf{x})$\\
      \midrule
      $1$ & $2.0$ & $a_{1} \cdot S$ \\
      $2$ & $1.0$ & $a_{2} \cdot S$ \\
      $3$ & $0.5$ & $a_{3}$ \\
      \bottomrule
    \end{tabular}
    \caption{Summary of each reaction channel in Kendall's process.
    The initial abundance of species $S$ is set to $5$ particles.
    \label{tab:si-table1}}
  \end{table}
  
\begin{model}Michaelis-Menten enzyme kinetics\end{model}
  \begin{equation*}
    \left\{
    \begin{aligned}
      S + E & \rightarrow SE &\text{Binding}\\
      SE    & \rightarrow S + E &\text{Dissociation}\\
      SE    & \rightarrow P + E
      &\text{Conversion}
    \end{aligned}
    \right.
  \end{equation*}

  \begin{table}[!htbp]
    \centering
    \ra{1.3}
    \begin{tabular}{@{}crr@{}}
      \toprule
      Reaction $j$ & Rate constant $a_{j}$ & Propensity $r_{j}(\mathbf{x})$\\
      \midrule
      $1$ & $0.00166$ & $a_{1} \cdot S \cdot E$ \\
      $2$ & $0.0001$  & $a_{2} \cdot SE$ \\
      $3$ & $0.1$     & $a_{3} \cdot SE$ \\
      \bottomrule
    \end{tabular}
    \caption{Summary of each reaction channel in the Michaelis-Menten model.
    This model is simulated with initial molecule counts $(S, E, SE, P) = (301, 120, 0, 0)$.
    \label{tab:si-table2}}
  \end{table}
  
\begin{model}Auto-regulatory genetic network\end{model}
  \begin{equation*}
    \left\{
    \begin{aligned}
      gene             & \rightarrow gene + RNA &\text{Transcription}\\
      RNA              & \rightarrow RNA + P &\text{Translation}\\
      P + P            & \rightarrow P_{2}   &\text{Dimerization}\\
      P_{2}            & \rightarrow P + P   &\text{Dissociation}\\
      gene + P_{2}     & \rightarrow P_{2}gene &\text{Repression binding}\\
      P_{2}gene        & \rightarrow P_{2} + gene &\text{Reverse repression binding}\\
      RNA              & \rightarrow \emptyset &\text{RNA degradation}\\
      P                & \rightarrow \emptyset &\text{Protein degradation}
    \end{aligned}
    \right.
  \end{equation*}

  \begin{table}[!htbp]
    \centering
    \ra{1.3}
    \begin{tabular}{@{}crr@{}}
      \toprule
      Reaction $j$ & Rate constant $a_{j}$ & Propensity $r_{j}(\mathbf{x})$\\
      \midrule
      $1$ & $0.01$ & $a_{1} \cdot gene$ \\
      $2$ & $10.0$ & $a_{2} \cdot RNA$ \\
      $3$ & $1.0$  & $a_{3} \cdot P \cdot (P - 1)$ \\
      $4$ & $1.0$  & $a_{4} \cdot P_{2}$ \\
      $5$ & $1.0$  & $a_{5} \cdot gene \cdot P_{2}$ \\
      $6$ & $10.0$ & $a_{6} \cdot P_{2}gene$ \\
      $7$ & $0.1$  & $a_{7} \cdot RNA$ \\
      $8$ & $0.01$ & $a_{8} \cdot P$ \\
      \bottomrule
    \end{tabular}
    \caption{Summary of each reaction channel in the negative auto-regulation gene network.
    The gene network is initialized with a few copies of the underlying gene.
    Specifically, we set $(gene, RNA, P, P_{2}, P_{2}gene) = (10, 0, 0, 0, 0)$.
    \label{tab:si-table3}}
  \end{table}
  
\begin{model}Dimerization-decay\end{model}
  \begin{equation*}
    \left\{
    \begin{aligned}
      S_{1}   & \rightarrow \emptyset &\text{decay} \\
      2 S_{1} & \rightarrow S_{2}     &\text{dimerization}\\
      S_{2}   & \rightarrow 2 S_{1}   &\text{dissociation}\\
      S_{2}   & \rightarrow S_{3}     &\text{conversion}
    \end{aligned}
    \right.
  \end{equation*}

  \begin{table}[!htbp]
    \centering
    \ra{1.3}
    \begin{tabular}{@{}crr@{}}
      \toprule
      Reaction $j$ & Rate constant $a_{j}$ & Propensity $r_{j}(\mathbf{x})$\\
      \midrule
      $1$ & $1.0$   & $a_{1} \cdot S_{1}$ \\
      $2$ & $0.002$ & $a_{2} \cdot S_{1} \cdot (S_{1} - 1)$ \\
      $3$ & $0.5$   & $a_{3} \cdot S_{2}$ \\
      $4$ & $0.04$  & $a_{4} \cdot S_{2}$ \\
      \bottomrule
    \end{tabular}
    \caption{Summary of each reaction channel in the dimerization-decay model.
    We initialize simulations of this process with $(S_{1}, S_{2}, S_{3}) = (1000, 0, 0)$.\label{tab:si-table4}}
  \end{table}
  
\begin{model}Yeast polarization\end{model}
  \begin{equation*}
    \left\{
    \begin{aligned}
      \emptyset      & \rightarrow R &\text{receptor upregulation} \\
      R              & \rightarrow \emptyset &\text{receptor downregulation} \\
      L + R          & \rightarrow RL + L &\text{ligand binding} \\
      RL             & \rightarrow R &\text{ligand degradation} \\
      RL + G         & \rightarrow G_{a} + G_{bg} + RL &\text{protein activation} \\
      G_{a}          & \rightarrow G_{d} &\text{dephosphorylation} \\
      G_{d} + G_{bg} & \rightarrow G     &\text{rebinding} \\
      \emptyset      & \rightarrow RL    &\text{bound receptor upregulation}
    \end{aligned}
    \right.
  \end{equation*}

  \begin{table}[!htbp]
    \centering
    \ra{1.3}
    \begin{tabular}{@{}crr@{}}
      \toprule
      Reaction $j$ & Rate constant $a_{j}$ & Propensity $r_{j}(\mathbf{x})$\\
      \midrule
      $1$ & $0.0038$  & $a_{1}$ \\
      $2$ & $0.0004$  & $a_{2} \cdot R$ \\
      $3$ & $0.42$    & $a_{3} \cdot L \cdot R$ \\
      $4$ & $0.01$    & $a_{4} \cdot RL$ \\
      $5$ & $0.011$   & $a_{5} \cdot RL \cdot G$ \\
      $6$ & $0.1$     & $a_{6} \cdot G_{a}$ \\
      $7$ & $0.00105$ & $a_{7} \cdot G_{d} \cdot G_{bg}$ \\
      $8$ & $3.21$    & $a_{8}$ \\
      \bottomrule
    \end{tabular}
    \caption{Summary of each reaction channel in the simplified model of pheromone-induced G-protein cycle in \textit{Saccharomyces cerevisiae}.
    Initial molecule counts are set to $(R, L, RL, G, G_{a}, G_{bg}, G_{d}) = (50, 2, 0, 50, 0, 0, 0)$.
    \label{tab:si-table5}}
  \end{table}

\subsection{Benchmark parameters}

Our benchmark results measure how quickly a particular software tool generates $10$ realizations of a given stochastic process up to a predetermined time $t_{\mathrm{final}}$; let us call this the \textit{simulation task}.
Specifically, we test the performance of \textit{similar implementations} of Gillespie's direct method across \pkg{StochPy}, \pkg{StochKit}, \pkg{Gillespie.jl}, and \pkg{BioSimulator.jl} based on our simulation task.
We summarize the parameters to the simulation task in Table~\ref{tab:si-table6}.
The results reported in Table 2 and Table 4 reflect statistics dervied from $1000$ samples of the simulation task.

\begin{table}[!htbp]
  \centering
  \ra{1.3}
  \begin{tabular}{@{}rrrrr@{}}
    \toprule
              & \multicolumn{1}{c}{$t_{\mathrm{final}}$} & \multicolumn{1}{c}{$n_{\mathrm{saves}}$} & \multicolumn{1}{c}{seed} & \multicolumn{1}{c}{$\epsilon$} \\
    \midrule
    Model $1$ &   $4$ & $1000$ & $5357$ & $0.03$ \\
    Model $2$ &  $50$ & $1000$ & $5357$ & $0.03$ \\
    Model $3$ & $500$ & $1000$ & $5357$ & $0.03$ \\
    Model $4$ & $100$ & $1000$ & $5357$ & $0.03$ \\
    Model $5$ & $100$ & $1000$ & $5357$ & $0.03$ \\
    \bottomrule
  \end{tabular}
  \caption{Summary of each parameter used in simulating each benchmark model.
  Here $t_{\mathrm{final}}$ denotes the end time of a simulation trajectory.
  The value $n_{\mathrm{saves}}$ denotes the number of time points a simulator should save under fixed-interval output.
  The $\epsilon$ value shown here is used only for both $\tau$-leaping methods in Table 2.
  Lastly, the \textit{seed} parameter is used to initialize a random number generator prior to each simulation task.
  This ensures that a simulator generates the same $10$ trajectories for each sample measurement.
  \label{tab:si-table6}}
\end{table}

\subsection{Additional notes}

\pkg{StochKit2}, \pkg{Gillespie.jl}, and \pkg{BioSimulator.jl} were benchmarked using the \code{@benchmark} macro from \proglang{Julia}'s \pkg{BenchmarkTools.jl} package.
This allowed the authors to set the number of samples to collect and enforce a time limit for each benchmark.
Since \pkg{StochKit} is a command-line tool, it was invoked using \proglang{Julia}'s \code(run) command to launch a subprocess.
Our results account for the overhead introduced in starting a new process by comparing against simulation times reported by \pkg{StochKit} at the command-line.
\pkg{StochPy} tracks a simulation's run time and makes it accesible in a scripting environment, so we deferred to these timing results.
The summary statistics reported in Table 2 and Table 4 were generated in \proglang{Julia}.

\pkg{Gillespie.jl} does not support aggregating multiple simulation runs by default.
We implemented a simple wrapper for the package's \code{ssa} command to mimic the behavior of the other software tools:
\begin{minted}[fontsize=\scriptsize,samepage]{julia}
function ssa_wrapper(x, x0, F, nu, parms, t_final, n_trial)
  for i in 1:n_trial
    # reset x to its initial value x0
    copy!(x, x0)

    # run the simulation
    ssa(x, F, nu, parms, t_final)
  end

  return nothing
end
\end{minted}